\documentclass{sig-alternate}
\usepackage{graphicx}

\usepackage{amsmath,amssymb}
\usepackage{subfigure}
\usepackage{color}
\usepackage{url}

\newcommand{\reminder}[1]{{\textcolor{red}{\textbf{}}}}

\begin{document}

\title{Coevolution of Network Structure and Content}

\numberofauthors{5}
\author{
\alignauthor Chun-Yuen Teng\\
\affaddr{School of Information}\\
\affaddr{University of Michigan}\\
\affaddr{Ann Arbor, MI 48109}\\
\email{chunyuen@umich.edu}
\alignauthor  Liuling Gong \\
\affaddr{School of Information}\\
\affaddr{University of Michigan}\\
\affaddr{Ann Arbor, MI 48109}\\
\email{llgong@umich.edu}
\alignauthor Avishay Livne
\affaddr{EECS}\\
\affaddr{University of Michigan}\\
\affaddr{Ann Arbor, MI 48109}\\
\email{avishay@umich.edu}
\and
\alignauthor Celso Brunetti \\
\affaddr{Carey Business School} \\
\affaddr{Johns Hopkins}\\
\affaddr{Baltimore, MD 21202}\\
\email{celsob@jhu.edu}
\alignauthor Lada Adamic\\
\affaddr{School of Information}\\
\affaddr{University of Michigan}\\
\affaddr{Ann Arbor, MI 48109}\\
\email{ladamic@umich.edu}
}

\maketitle

\begin{abstract}
As individuals communicate, their exchanges form a dynamic network. We demonstrate, using time series analysis of communication in three online settings, that network structure alone can be highly revealing of the diversity and novelty of the information being communicated. Our approach uses both standard and novel network metrics to characterize how unexpected a network configuration is, and to capture a network's ability to conduct information. We find that networks with a higher conductance in link structure exhibit higher information entropy, while unexpected network configurations can be tied to information novelty. We use a simulation model to explain the observed correspondence between the evolution of a network's structure and the information it carries.

\end{abstract}


\section{Introduction}
The Web is a highly dynamic network, and has been made all the more so with the increased use of social media. As networked individuals communicate online, their interactions leave dynamic network traces.  However, models of information flow assume a static network over which information can be transmitted, whether they are modeling the adoption of ideas and behavior~\cite{Watts30042002,centola2010spread,aral2010creating}, convergence of opinion~\cite{bindel2011bad}, or the speed and extent of innovation~\cite{Lazer01122007}.

In practice, networks are rarely static, unless one considers only the strongest and most stable ties~\cite{doi:10.1056/NEJMsa066082} or experimentally dictates the network topology to be fixed~\cite{centola2010spread}.  However, even stable ties transfer information at different rates~\cite{shelley1990information,kossinets2008structure,Onnela01052007}, and a portion of information flow occurs outside of established social ties~\cite{bakshy2009social}. New ties are also induced by information flow, e.g. a Pakistani Twitter user who inadvertently live-tweeted the Bin Laden assassination quickly gained tens of thousands of new followers on Twitter. This points to a need to
approach the relationship between network structure and information content in a substantively different way. 

In this paper, rather than treating the network structure as static, we specifically use its dynamic nature to infer two properties of the information being communicated through the network. 
The first is the diversity of the information; whether everyone is talking about the same topic or whether one is observing many disparate conversation topics being discussed. The second is the novelty of the information; whether individuals in the network are continuing to talk about the same topic they talked about in the previous time period, or whether new topics have arisen that are different from what has been discussed before. For example, one could imagine oneself at a dinner party, where most conversations are out of earshot, but one can easily observe who is conversing with whom. While individuals are milling about and chatting casually, one might expect that the information being exchanged corresponds to a diverse set of topics. If individuals with no common history start communicating, or if several people communicate with a single individual in close succession, one might expect an external event, such as the arrival of new information, to be the driving force of what is being communicated.

The paper is organized as follows. After describing related work in Section~\ref{sec:related} and data sets in Section~\ref{sec:data}, we present our approach, including network and content variables and their characteristics, in Section~\ref{sec:methods}. In Section~\ref{sec:results} we demonstrate that network variables can be highly predictive of content characteristics, and delve into the correspondence between individual network and content variables. In Section~\ref{sec:simulation} we present a simulation model of networked communication that can reproduce the observed patterns, and discuss future directions in Section~\ref{sec:conclusion}.

\subsection{Related Work}\label{sec:related}
The dynamic nature of web content has been of interest because of its implications for search and retrieval~\cite{Ntoulas04whatsnew,adar2009web}. In particular, changes in content and link structure can be used to find trending content. Sarma et al.~\cite{sarma2011dynamic} used bursts in appearance of connections between entities in text to detect events. Although they related network structure to properties of content, the network was generated from entities within the content itself. Lin et al.~\cite{lin2010pet} analyzed a joint model of network and topic evolution to track and predict topic popularity, but did not explicitly examine network structure. Two other studies used aggregate volume of interactions in social media to infer the evolution of content, but did not explicitly examine the networks' structure. Mathioudakis et al.~\cite{mathioudakis2010early} used the volume of interactions between nodes in social media, along with other variables, to identify attention gathering items early on in their lifecycle. In a non-Web context, Saavedra et al.~\cite{saavedra2011synchronicity} showed that the communication volume among stock traders correlated with synchrony in their trading behavior and correspondingly with profits.   

Prior work that has examined time-evolving network structure explicitly has shown that changes in a network's structure can be reflective of events and trends.  In the Graphscope project~\cite{sun2007graphscope}, changes in community structure of email communication networks were tied to events within a company. Leicht et al.~\cite{leicht07cite} showed that breaks in supreme court citation patterns corresponded to changes in the court's ideology. Adamic et al.~\cite{adamic10} correlated time series of networks of traders trading in commodity futures contracts, with financial variables related to the trades, such as returns, volatility and duration. However, the notion of information entering into the market was implicit, in the sense that the prices and quantities of contracts traded reflected the information the traders held about the future value of the contract. In contrast, in this paper we explicitly analyze the information that is being directly communicated with each activation of an edge.

Our approach is to segment the network and content by time, measure network and content properties for each segment, and correlate their time series.

\section{Data}\label{sec:data}
To demonstrate the flexibility of the approach, we employ three sets of network data representing information exchange. The first data set is drawn from Twitter, a microblogging platform that allows users to post 140 character messages. Twitter messages can contain references to other users, and these references comprise the edges of the network. As Twitter is a network of hundreds of millions of nodes around the globe, we focus on smaller communities within the network that can be expected to be communicating about similar content. 

We selected 9 accounts that were tagged with ``researcher'' and had at least 200 followers (other accounts that subscribe to the content tweeted by the researcher). We then gathered the set of accounts followed by each researcher, with sets varying in size from 257 to 1342. We constructed 9 corresponding networks of mentions and retweets of one account by another. In order to focus on just specific communities that are followed by each of the researchers, we applied a community finding algorithm to each of the 9 networks~\cite{Newman2006}, filtering out any community of size less than ten. As a result, we obtain the time-evolving network for filtered sets of Twitter accounts followed by a researcher, with size ranging from 164 to 982 and associated tweets numbering from 92,195 to 346,399. To construct the time series, we segmented the data for each network into 800 tweets, and calculate network and content variables for each segment. The resultant time series range in length between communities from 116 to 443 segments, with more active communities generating longer time series.

The second data set is drawn from the virtual world Second Life (SL), where almost all of the content, in the form of assets, is created and distributed by the users themselves. Here we are observing exchanges of discrete packages of information between July 2008 to December 2009, each one being either a landmark (a bookmark of a location within SecondLife), or a gesture (a script that allows an avatar to execute a particular motion or to utter a sound). A typical landmark might correspond to the location of a storefront or club, while a gesture might correspond to a dance move or a laugh.  
An in-depth analysis of this data is presented in~\cite{bakshy2009social}. In this context, content diversity corresponds to many different landmarks and gestures being transferred between users. Novelty is high when what is transferred during one time segment is different from what was being transferred in a previous segment.

Groups in SL are used primarily to exchange information and coordinate activities, but also sometimes to share land ownership and other functions. We selected the 100 groups with
the highest number of asset transfers occurring within each group. For each group, we divide the data into segments of 50 transfers each, with nodes in the network corresponding to users, and edges corresponding to one or more asset transfers between two users in the same group. In contrast to the Twitter data, where any user actively tweeting is included as a node in the segment, a SecondLife user is included only if engaging in an asset transfer with another member of a group.

The third data set is constructed from the publicly available Enron email data~\cite{klimt2004enron}, with all nodes treated as belonging to a single community. We limit our analysis to the 155 most active email accounts, with each individual both sending and receiving at least 50 emails in the 2000-2001 time period. Emails sent to $> 20$ individuals were interpreted as mass mailings and removed in a pre-processing step. We performed the analysis with the entire text of the emails, excluding attachments, and then repeated the analysis with quoted text (from a reply or a forwarded message) removed. 

\section{Methods} \label{sec:methods}
\subsection{Characterizing the network's structure}
The interval length used to segment a network can affect the measured properties~\cite{clauset2007persistence}. The number of nodes and edges typically increases the longer the interval. Longer intervals are also likely to smooth out daily or weekly periodicities that may be picked up with shorter segments. To control for these daily and weekly cycles, as well as drift in overall activity, we segment the networks by a fixed number of actions, rather than absolute time. Twitter networks are segmented into 800 tweets each, maximizing the number of edges per segment while yielding a sufficient number of data points in the series. The higher number of actions is selected because only 
2\% of tweets contain mentions of other users {\em within the same group}, even though roughly 47\% of tweets 
for our selected users do mention other users.  The SL networks are segmented by 50 asset transfers, but the results we report in the following sections are qualitatively similar when data is segmented by either 25 or 75 transfers. The Enron email data is segmented into networks formed by 100 consecutive emails, with qualitatively similar results for segments of 50 and 200 emails.

To each network we applied the same set of standard network metrics, including number of nodes, number of edges, reciprocity (i.e. the proportion of edges that are bidirectional), clustering coefficient (i.e. the number of connected triples in the network that form a closed triad), centralization (i.e. the gini coefficient of the undirected degree distribution), correlations of degrees of nodes across each edge, average degree, standard deviation of degree, and the size of largest strongly and weakly connected components. 

To complement these standard network metrics, we employ two additional measures. The first, cycle-free effective conductance, defined in ~\cite{koren06proximity}, captures the potential for directed information flow between a pair of nodes $i$ and $j$ in the network:    \begin{equation}
        C_{ij} = \sum_p\prod _{(k,l)}\frac{w(k,l)}{deg(k)}
      \end{equation}
where $p$ is a path between nodes $i$ and $j$, $k$ and $l$ are nodes in the path with edge $(k,l)$ of weight $w(k,l)$ between them, and $deg(k)$ is the out-degree of node $k$. From this definition, there will be high conductance from node $i$ to node $j$ if there are many paths, if the paths are short, and if they go through nodes that have few neighbors besides the ones on the path. To obtain the conductance of the entire graph, we sum over all pairs of nodes present in the time segment:
$
        C = \sum_{ij}{C_{ij}}.
$

By capturing the amount of connectivity within the network, conductance represents the capacity of the network to transmit information within a given time segment.

\begin{figure}[tbh]
\centering
\includegraphics[width=0.9\columnwidth]{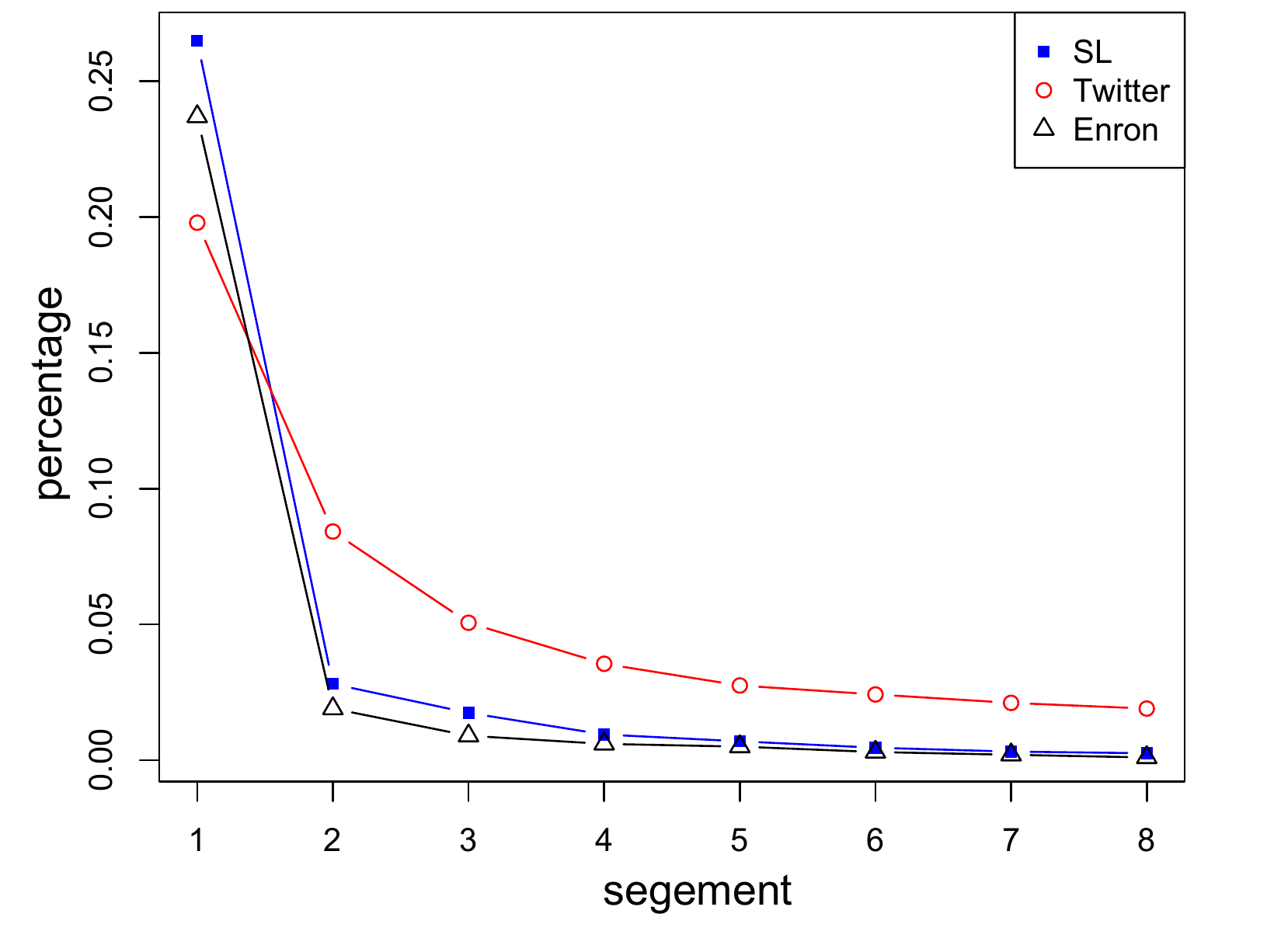}
\caption{The likelihood that an edge is repeated as a function of lag.\label{fig:lifespan}}
\end{figure}

In addition to metrics characterizing the network structure within one segment, we employed others to capture the change in structure between segments. As shown in Figure~\ref{fig:lifespan}, the three networks are highly dynamic, with the probability of an edge repeating diminishing the farther apart the network segments are. A simple measure of network stability is the overlap in edges between two consecutive time segments, as captured by the Jaccard coefficient: \begin{equation}
J_{E_{t},E_{t-1}} = \frac{|E_{t} \cap E_{t-1}|}{|E_{t} \cup E_{t-1}|}
\end{equation}

While repeat edges are unsurprising, other edges may also be expected to occur, depending on whether information had the potential to flow indirectly between them in the recent past. This potential for information to flow between endpoints of an edge $e$ is captured by the past conductance $C_{e}(t-1)$. Nodes that can be expected to communicate at time $t$, will likely have communicated directly in the past {\em or} were linked by many indirect paths, corresponding to a high $C_{e}(t-1)$. We construct a network-level measure of expectedness $X_{t}$, by averaging $C_{e}(t-1)$ over all edges $e \in E_{t}$ present at time $t$.
\begin{equation}
X_{t} = \frac{1}{|E_{t}|} \sum_{e\in E_{t}}C_{e}(t-1)
\end{equation}

\begin{figure}[htb]
\centering
\includegraphics[width=1\columnwidth]{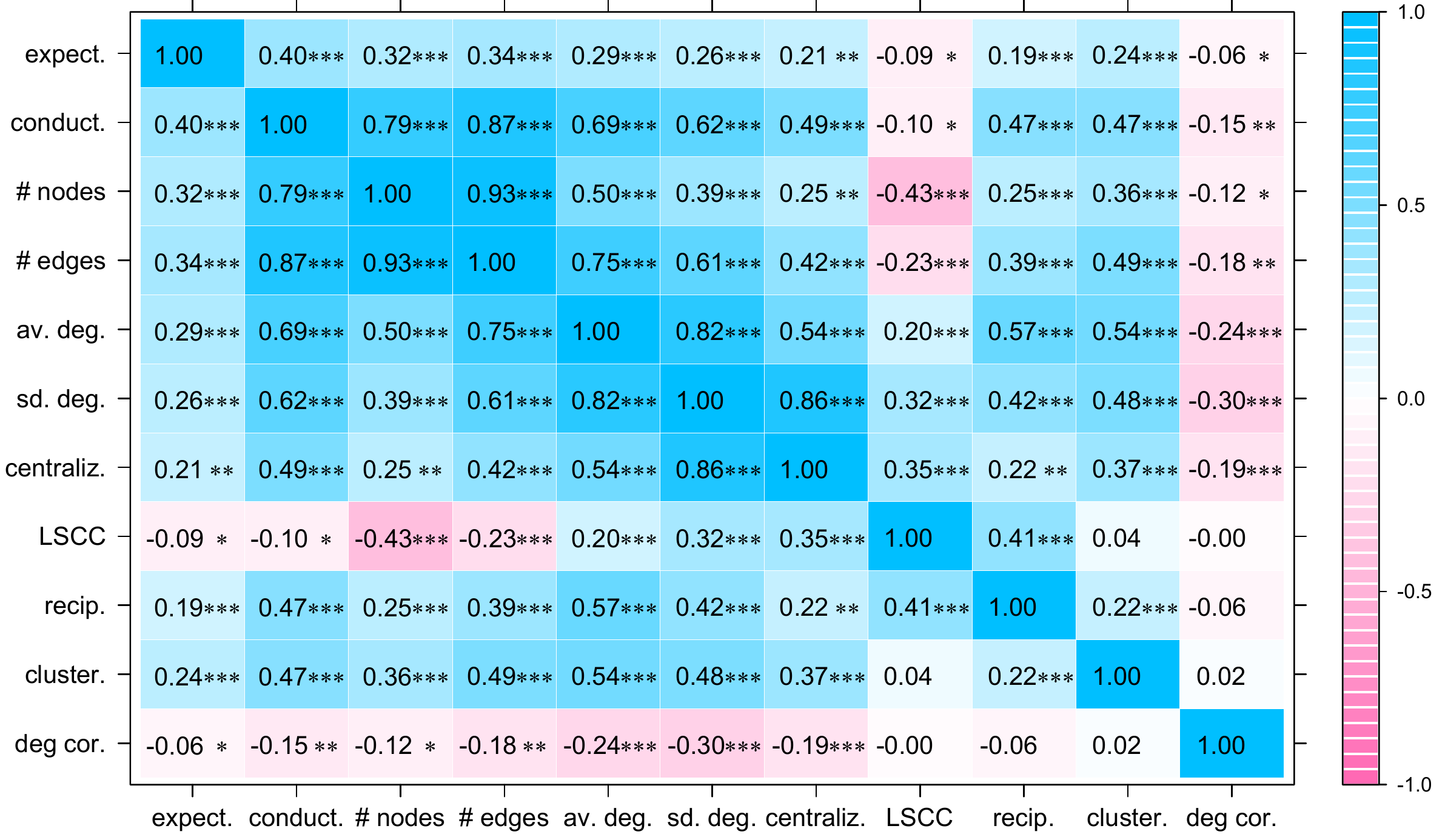}
\caption{Correlations between network metrics applied to communication within Twitter communities.\label{fig:netcorTwitter}}
\end{figure}

\begin{figure}[htb]
\centering
\includegraphics[width=1\columnwidth]{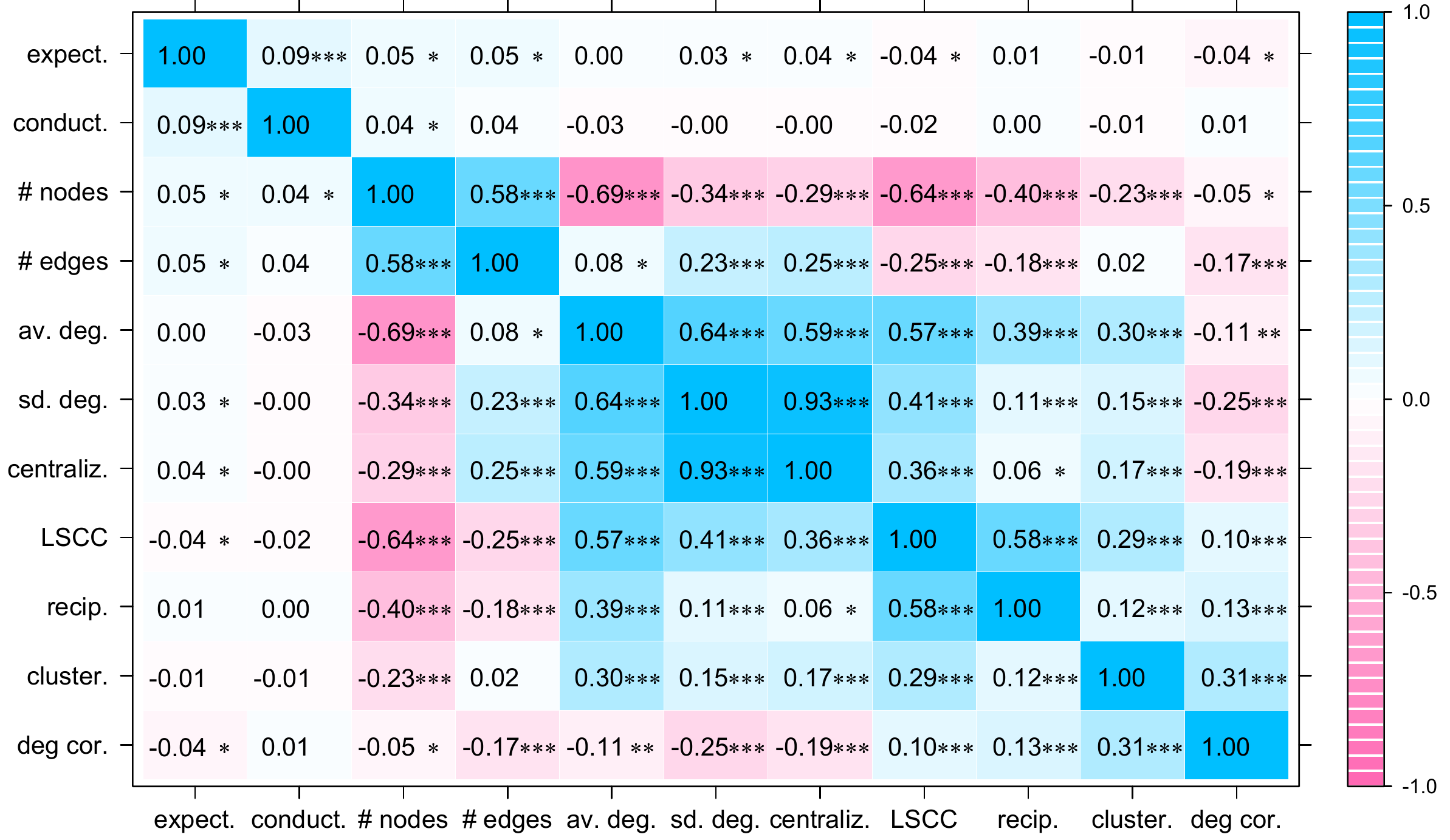}
\caption{Correlations between network metrics applied to communication within SecondLife.\label{fig:netcorSL}}
\end{figure}

\begin{figure}[htb]
\centering
\includegraphics[width=1\columnwidth]{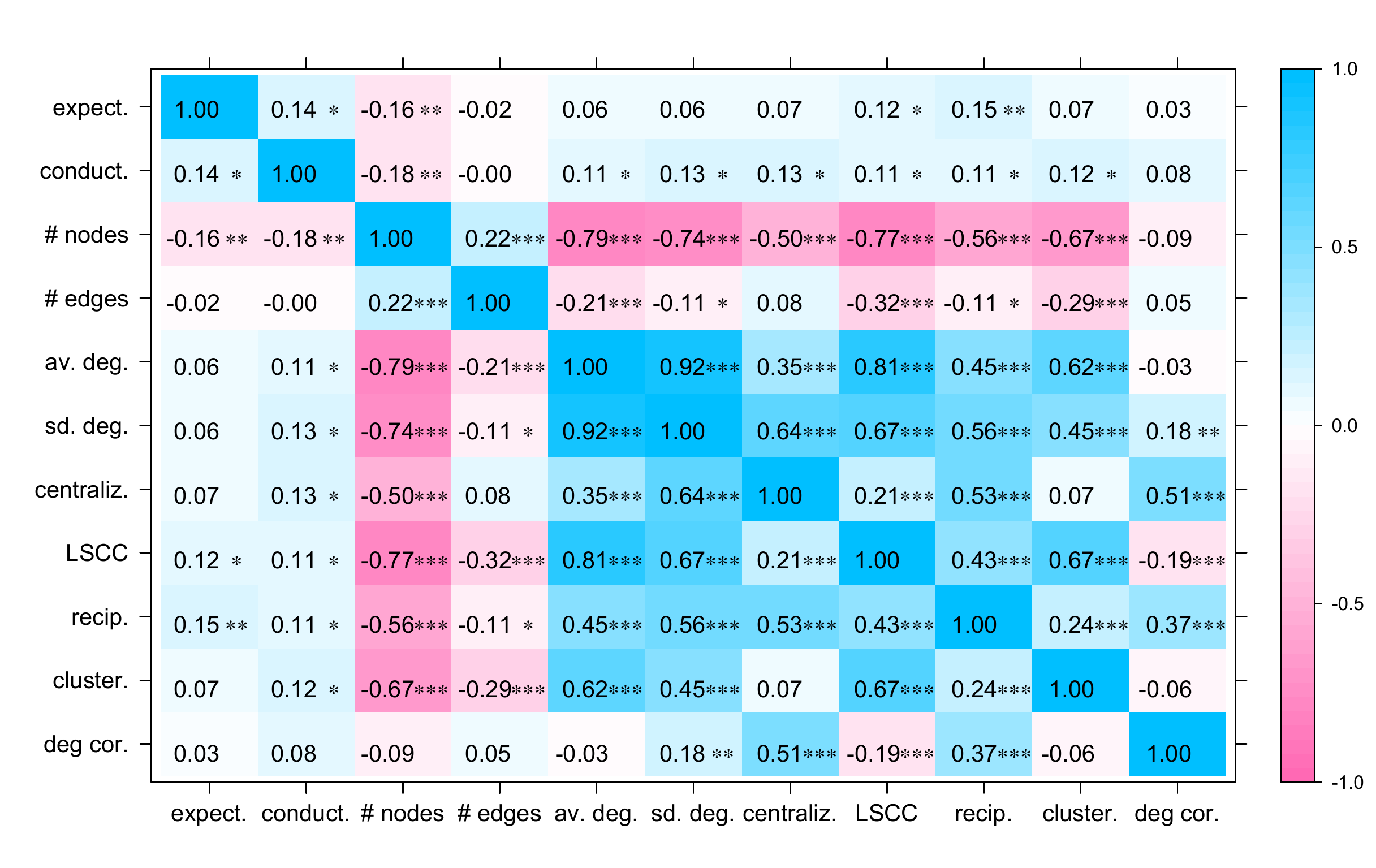}
\caption{Correlations between network metrics applied to Enron email correspondence.\label{fig:netcorEnron}}
\end{figure}

The network metrics are correlated amongst themselves, as shown in Figures~\ref{fig:netcorTwitter} (Twitter), \ref{fig:netcorSL} (SecondLife), and~\ref{fig:netcorEnron} (Enron).  Network segments with more nodes also tend to have more edges, as well as higher mean and variance of node degree. A large strongly connected component (LSCC), which potentially allows information to flow from any node that belongs to it to any other, is more likely to occur when there are more edges, reciprocity and clustering. However, when the edges are spread too thinly among a large number of nodes, the LSCC is smaller. Centralization, measuring the inequality in degrees among nodes, is positively correlated with standard deviation in degree. Conductance is uncorrelated with other measures in the SL and Enron data sets, and positively correlated with network size, e.g. the number of nodes and edges, in the Twitter communities. 

\subsection{Characterizing the network's content}
The above measures capture the changing shape of interactions. Additional measures allow us to characterize the content of those interactions. We select measures according to the type of content. In SecondLife, discrete assets are traded and their diversity is easily captured by the entropy $H_{A}$ of their distribution during a time segment. Whether the same or different assets are being traded relative to the previous time period is captured by the asset Jaccard $J_{A}$, which measures the overlap in the sets. To assess the similarity and novelty of the text in tweets and email messages, we transform a piece of text $x$ into a word frequency vector $\phi(x)$, scaled by the inverse document frequency of each word \cite{Salton1973}. The similarity of two texts $x$ and $y$ is given by a cosine similarity measure,
\begin{equation}
  \nonumber
  S(x, y) = \frac{\langle\phi(x), \phi(y)\rangle}{\|\phi(x)\| \cdot \|\phi(y)\|}.
\end{equation}

\begin{figure}[htb]
  \centering
  \includegraphics[width=0.9\columnwidth]{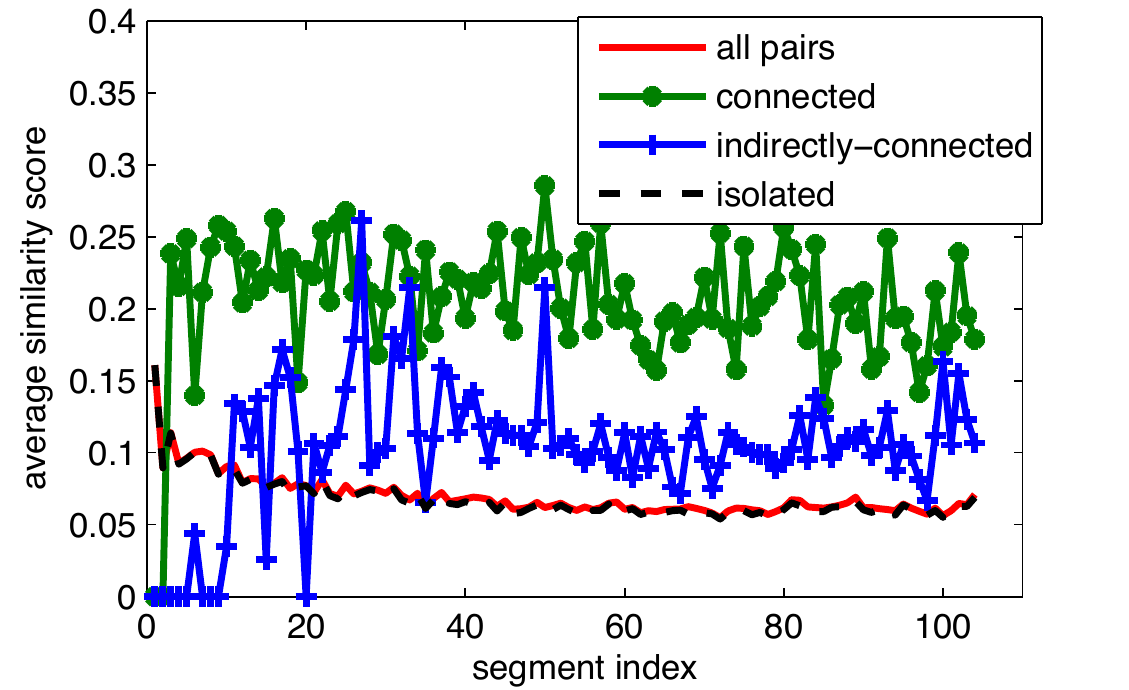}
  \caption{Time series of the average similarity score of the community of users followed by a  Twitter user.  \label{fig:ASS}}

\end{figure}
To measure the coherence of content exchange in the Enron email corpus, we calculate the pairwise similarity between all emails that comprise the network. In Twitter, as mentioned, content explicitly directed at other users within the same community comprises only a small fraction of all content broadcast by members of a community. To obtain a sufficient amount of content per segment, we aggregate posts by user, and evaluate similarity among user pairs. We differentiate between node pairs that share an edge, pairs that do not share an edge but are connected indirectly through other nodes, and pairs of nodes that are located in different components. Due to the low density of the Twitter graph segments, most nodes are located in separate components. Hence the average content similarity between any two nodes in the network is close to the similarity between nodes in different components, as shown in Fig. \ref{fig:ASS} for one sample Twitter community. 

Having captured the diversity of information shared during a single segment, we next quantify the extent to which content is changing  between time segments in the Enron and Twitter data sets. We adapt the technique described in \cite{Shmueli-Scheuer2010} to construct language models (LMs) corresponding to the aggregate content that was produced by all nodes in a given time segment. We then calculate the symmetric Kullback-Leibler (KL) divergence on the LMs of the two time segments to quantify the LM distance, defined as
\begin{eqnarray}
     \nonumber   D_{KL}^S(P_{t} \| P_{t-1}) & = &  \sum_{w \in W} P_{t} (w) \log \frac{P_{t} (w)}{P_{t-1} (w)} \\
        & & + P_{t-1} (w) \log \frac{P_{t-1} (w)}{P_{t} (w)},
\end{eqnarray}
where $P_t(w)$ is a distribution over vocabulary $w \in W$ at time $t$.  A high KL divergence implies that the content is novel, since it has changed substantially from the previous time segment. 

\subsection{Time series stationarity}
Our analysis entails finding contemporaneous correlations between time series, e.g. determining whether higher conductance corresponds to either higher or lower content diversity. Such analyses are sensitive to overall trends and seasonalities. For example, if the average degree in the network segments were to be increasing over time, and diversity were increasing as well, then one would measure a spurious positive correlation between the two time series. Therefore we need to establish that the series are stationary before proceeding with the analysis.

We test for stationarity using the augmented Dickey-Fuller (ADF) and the Phillips-Perron stationarity tests. For the ADF test, we select the optimal lag-length using the Bayesian information criterion; while for the Phillips-Perron test we adopt a Barlett kernel and Newey-West bandwidth selection. In the Twitter and Enron email data sets, all network and content variable time series were found to be stationary with p-values below 1 percent (therefore rejecting the null of a unit root). For the SecondLife dataset we excluded 46 groups which had one or more non-stationary time series. Results including these 46 groups are qualitatively similar to the results presented in the paper.

\section{Correspondence between \\network structure and content}\label{sec:results}
 
In aggregate, we find a strong correspondence between the evolving network structure of interactions, and the content that is being communicated. Before reporting the aggregate results, we take a brief detour examining a few sample segments in each setting to build intuition about their network and content characteristics.

    \begin{table}
    \caption{Sample group segments in SL}
    \begin{tabular}{ | l || c | c |}
    \hline \hline
    Parameter & Sample A & Sample B \\ \hline
     & \includegraphics[scale=0.15]{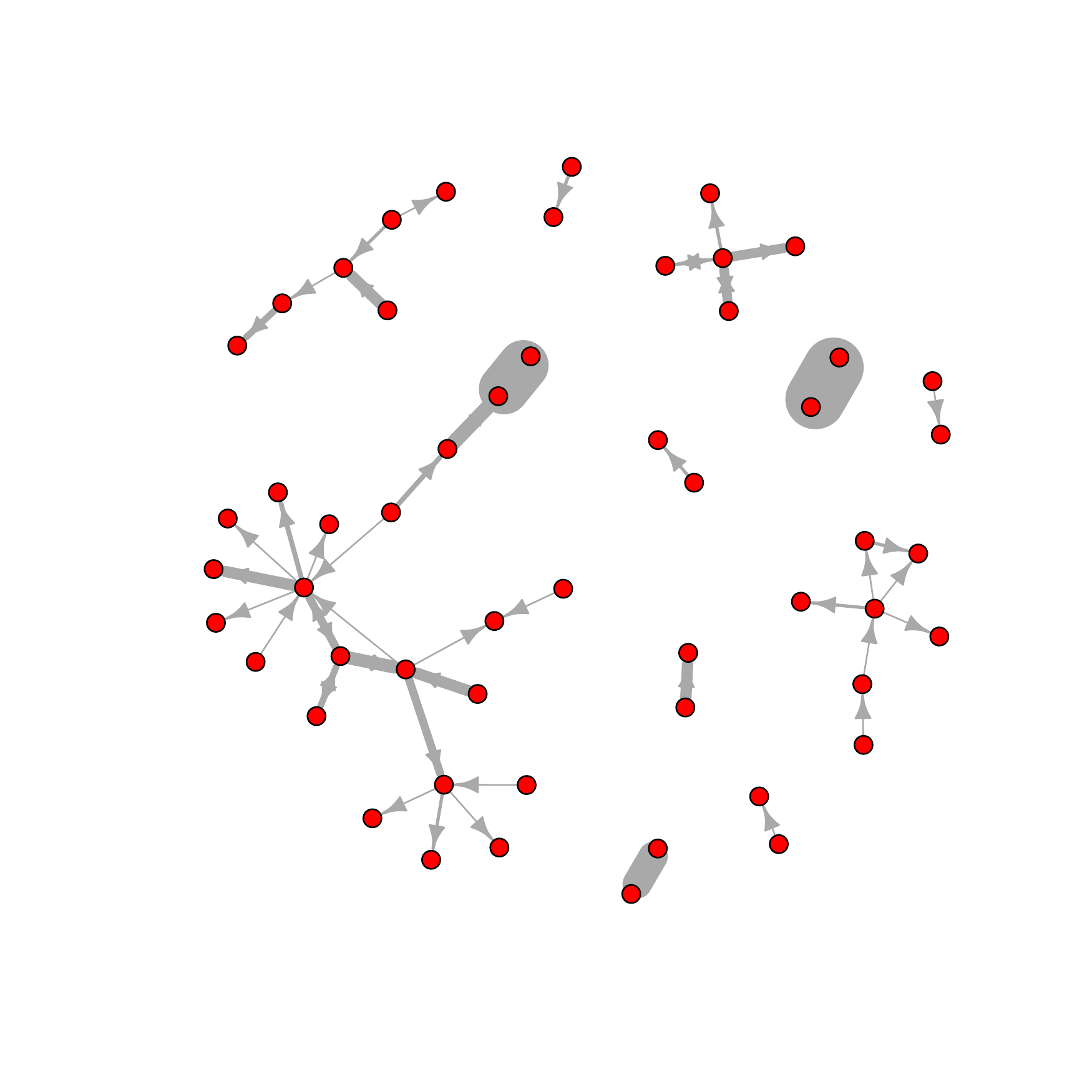} 
     &\includegraphics[scale=0.15]{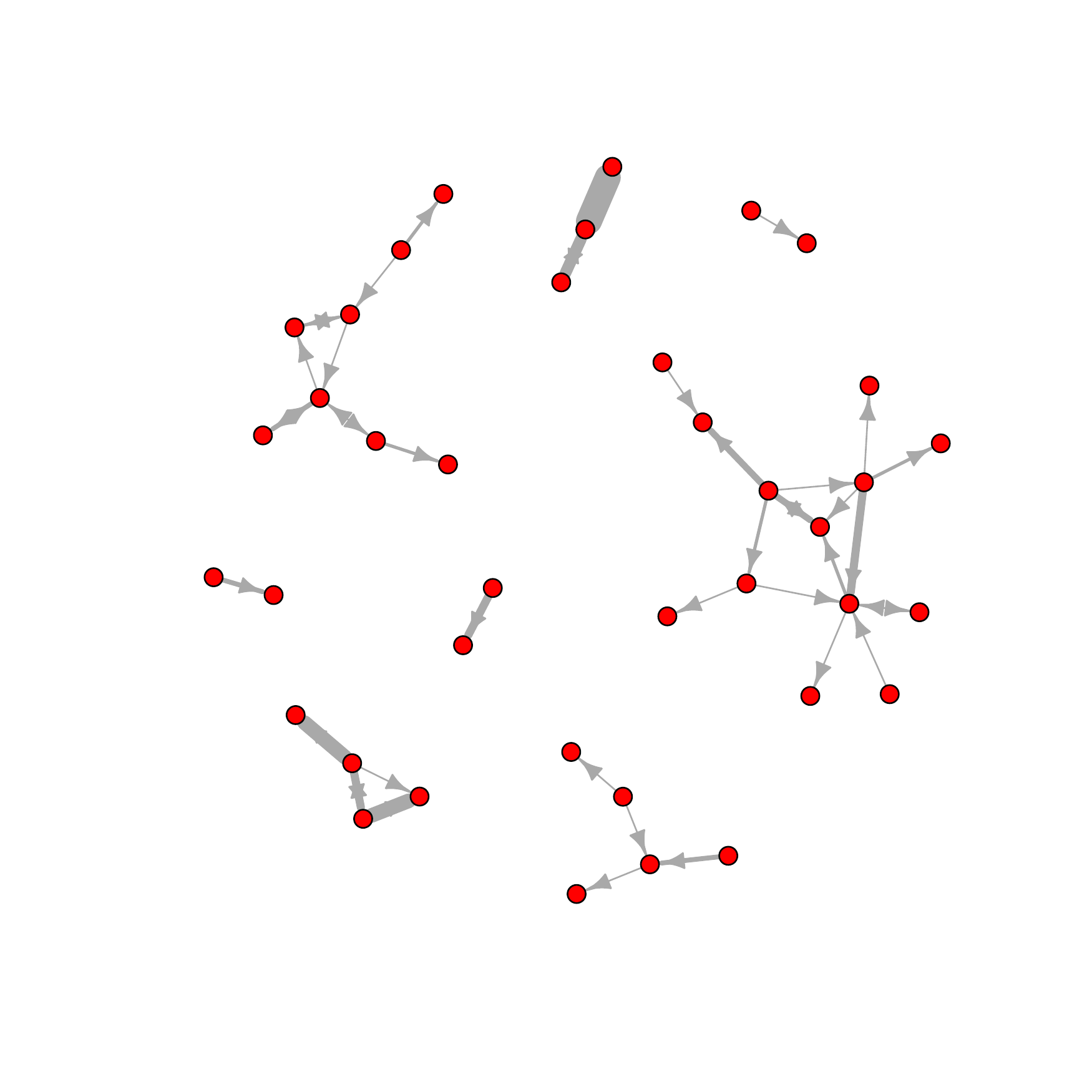} \\ \hline
   $C$ & 1088.36 &  128.95 \\ \hline
   $H_A$ &  7.84 &  3.84 \\ \hline
   $X$ & 54.56 & 16.09 \\ \hline
   $J_A$ & 0.18 & 0.06 \\ \hline \hline
    \end{tabular}
    \label{table:SLSample}
    \end{table}
    
\subsection{Sample illustrations of co-evolution of structure and content}
\subsubsection{Second Life}
We select a particular SL group of 296 users focusing on music, art, and fashion. Table ~\ref{table:SLSample} contrasts landmark exchange in two different sample segments. Segment A has higher network conductance relative to segment B. Correspondingly, the network snapshots reveal more numerous and larger connected components, as well as higher edge density. 

The segment with higher conductance also has higher asset entropy, $H_{A}$. Examining the top traded landmarks, shown in Table~\ref{table:SLSampleContentA}, we observe a diverse set of locations: a beach, an orientation space for new Spanish-speaking users, an adult entertainment locale, and a dance club. In contrast, segment B, exhibiting lower conductance and asset entropy, has a dominantly traded landmark, which, along with several others, is tied to fashion and shopping (see Table~\ref{table:SLSampleContentB}). 
   \begin{table}
    \caption{Top 5 assets in a SL group in segment A}
    \begin{tabular}{ | l | l |}
    \hline \hline
    Count & Description\\ \hline
7 & Tropical Waterfront\\  \hline
7  & Spanish Orientation\\  \hline
4  & Naughty Island \\  \hline
3  & Vienna / Wien\\  \hline
3  & The Garage Club \& Risa Island Dr \\  \hline \hline

    \end{tabular}
        \label{table:SLSampleContentA}
    \end{table}
    
    \begin{table}
    \caption{Top 5 assets in SL group in segment B}
    \begin{tabular}{ | l | l |}
    \hline \hline
    Count & Description\\ \hline
12  &Thea Tamura Fashion - Main Shop \\  \hline
5  & J's MainShop Tsukishima  \\  \hline
3  & Shop Misty, dresses,skin,tiny ca \\  \hline
3  & The Skylight Depa \\  \hline
3 &  Pensieri e P \\  \hline \hline

    \end{tabular}
    \label{table:SLSampleContentB}
    \end{table}

We can also contrast the novelty of the network structure and content between the two segments. The network configuration in A is more expected based on previous trades than the network configuration in B, giving A a higher expectedness value $X$. The continuity in network structure in A is accompanied by a higher continuity in content: the assets traded in A overlap more with the previous time step. One might speculate that a fashion-focused event drove both new exchanges and introduced new locations that captured a substantial portion of the attention in segment B.

\begin{table*}[htb]
\centering
\caption{Coevolution of network structure and content in a Twitter community as the Wikileaks story breaks. }
\begin{tabular}{|l ||c|c|c|c|c|}
 \hline \hline
Segment & 206 & 207 & 208 & 209 & 210\\ \hline
& \includegraphics[scale=0.13]{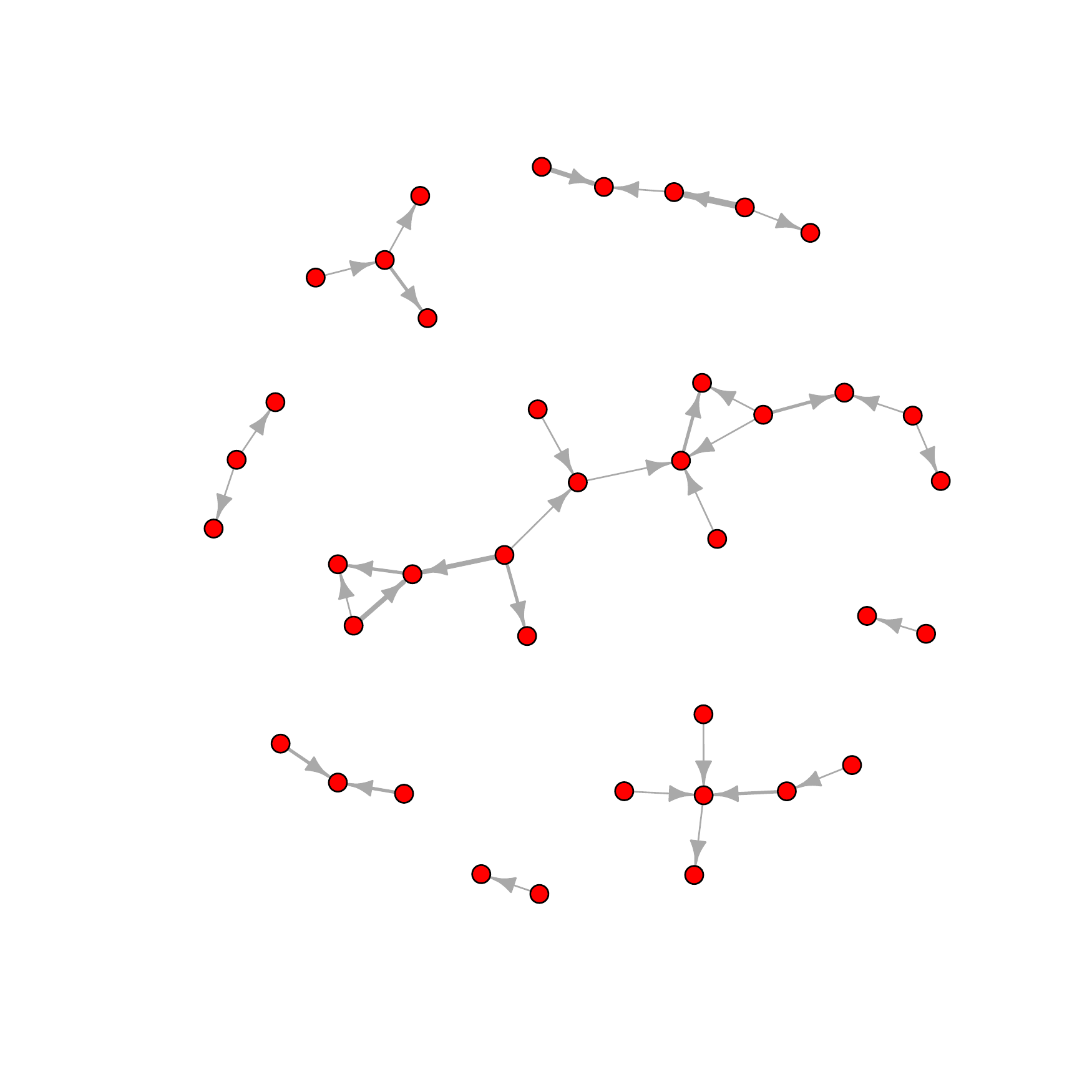} 
&\includegraphics[scale=0.13]{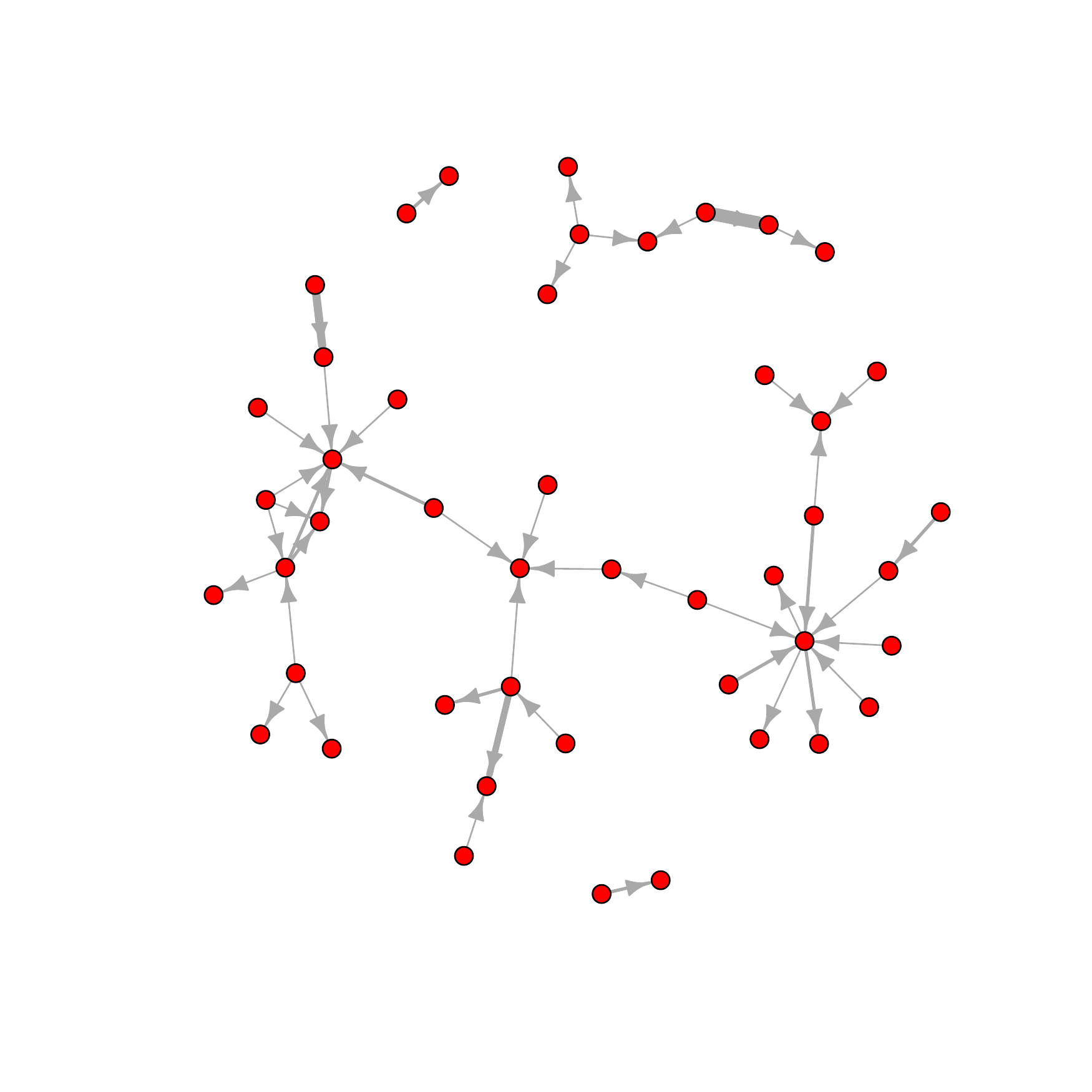} 
& \includegraphics[scale=0.13]{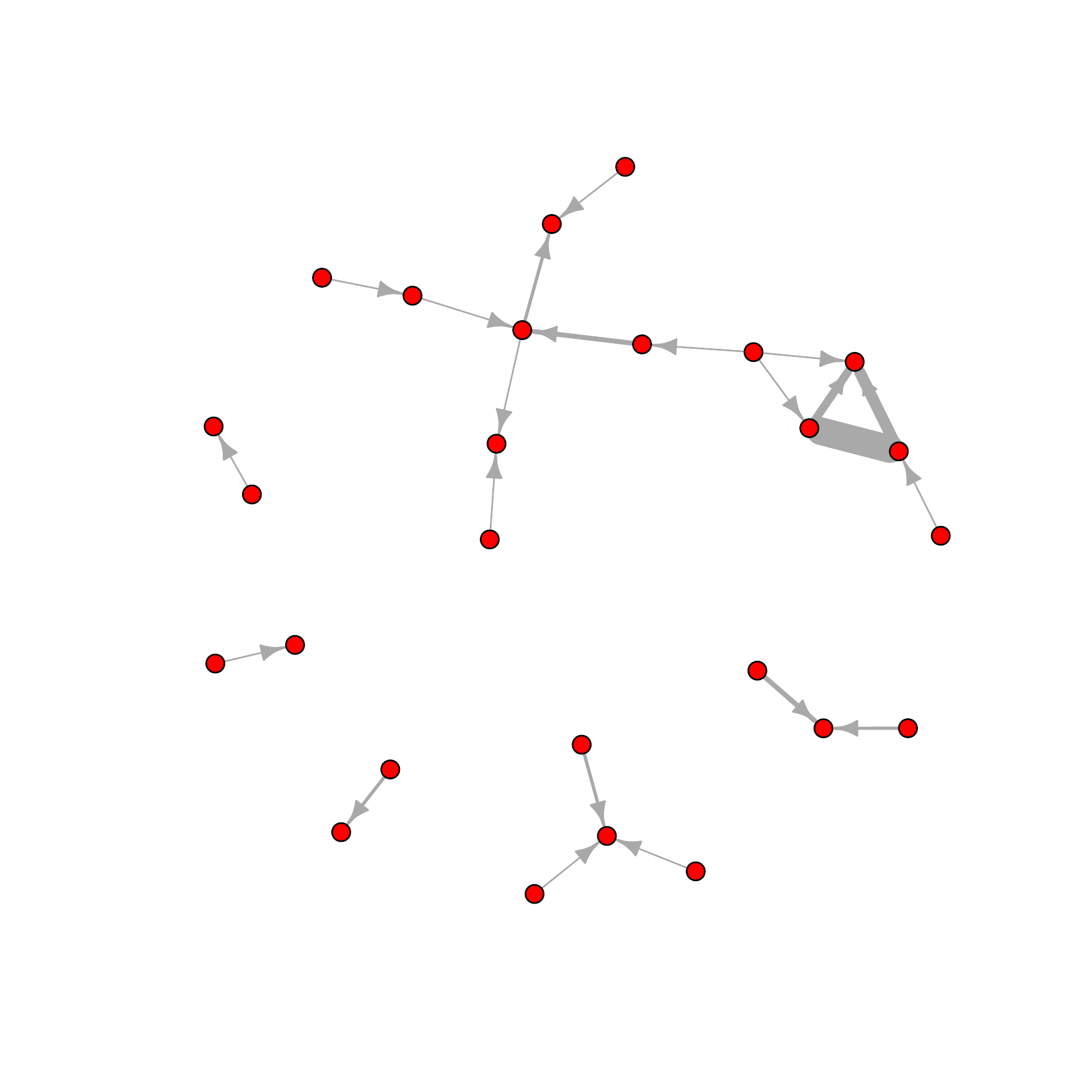} 
& \includegraphics[scale=0.13]{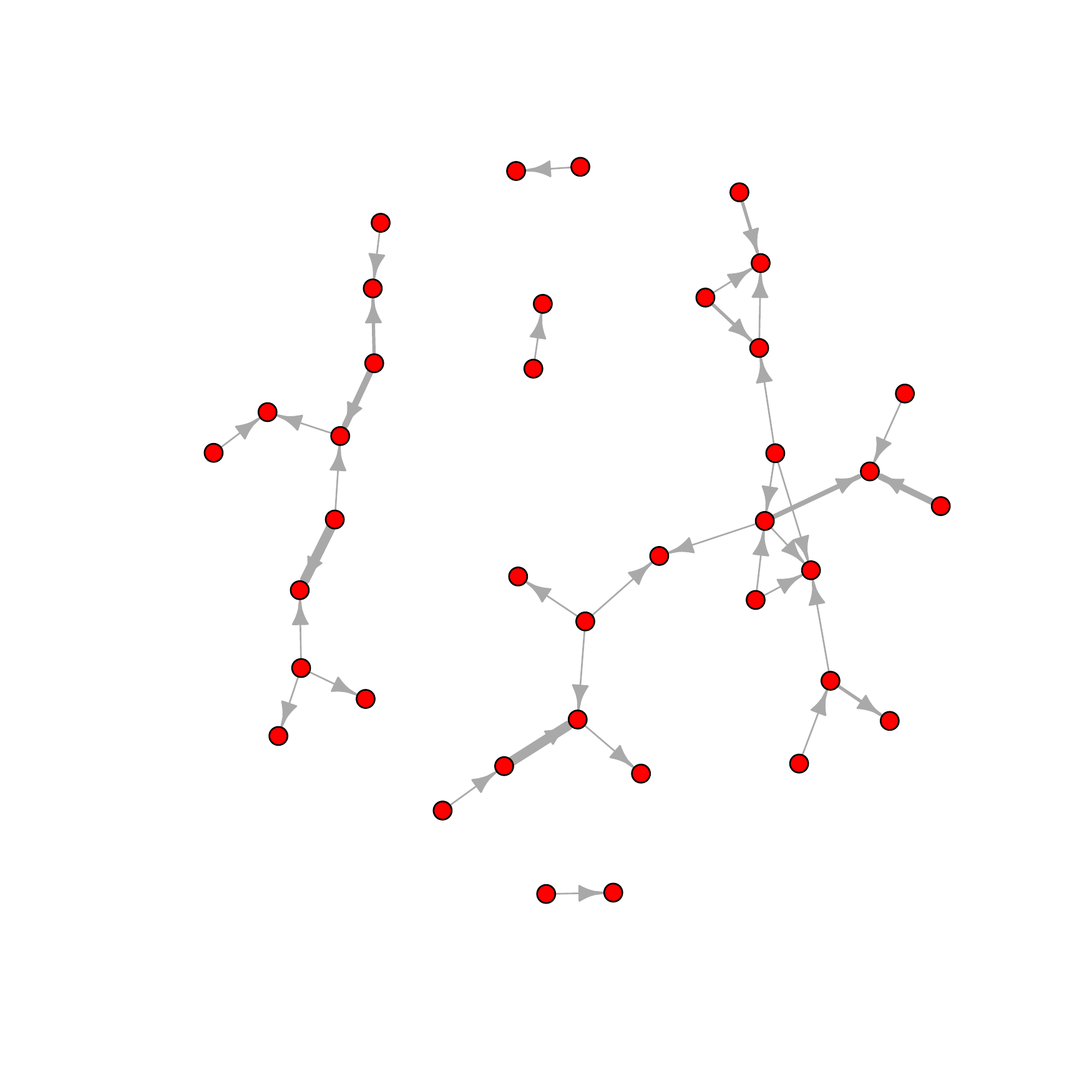} 
& \includegraphics[scale=0.13]{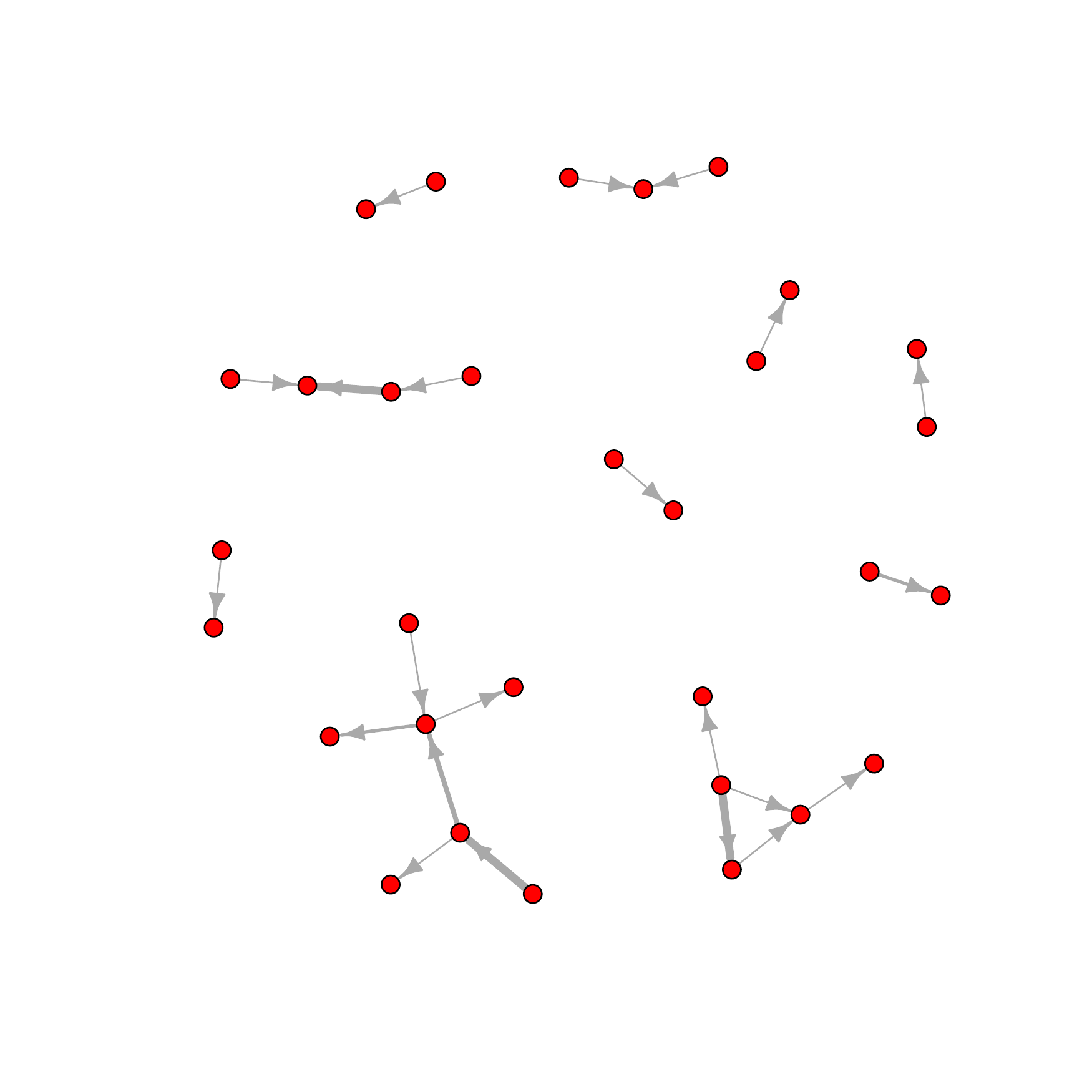}\\ \hline
conductance &  44.33 & 51.78 & 34.56 & 35.71 & 26.22\\ \hline
expectedness & 4.03 & 6.50 & 4.29 & 5.834 & 2.17 \\ \hline
similarity (unconn.) & 0.040 & 0.037 & 0.039 & 0.037 & 0.035\\ \hline
similarity (indir. conn.) & 0.22 & 0.04 & 0.18 & 0.25 & 0.04\\ \hline
LM dist. (all) & 14.00  & 12.96 & 8.79 & 12.47 & 16.03 \\ \hline
LM dist. (non-isolated) & 18.17 & 9.00 & 15.40 & 20.32 & 20.44 \\ \hline \hline
\end{tabular}
\label{table:Enron}
\end{table*}

\subsubsection{Twitter}
An example taken from the Twitter dataset is the outbreak of the Wikileaks story on 28 November 2010. In segment 207 we see a sudden appearance of many related terms ({\em wikileaks, cablegate, embassy, guardian}). This spread of new information is accompanied by an increase in expectedness and conductance scores compared to the preceding segment. The novelty and dominance of the story is less detectable in the content metrics. We see a moderate divergence in language models just for non-isolated nodes between segment 207 (in which the story broke) and segment 206. Once the story becomes dominant, there is high similarity between segments 207 and 208 across the network, suggesting that the community continued to discuss this trendy topic. In segment 210, when the topic diminishes and new terms start to dominate the discussion,
we see a steep decrease in both expectedness and conductance, whereas the language model divergence increases, indicating that new content is being discussed.

\subsubsection{Enron}
Two sample segments from Enron illustrate contrasting email network structure and content. Sample A, capturing a period shortly after the Enron scandal had unfolded, shows low content similarity and novelty. While there are some copies of press releases relating to the scandal, and an email of someone asking for a reference ``just in case something happens'', there are also many routine calendar meetings, updates on existing contracts, notices of a fax machine out of order, etc. The network of interactions has both high conductance and high expectedness.

In contrast, segment B contains cascading messages about a new contract with Dominion gas supply for the city of Tallahassee. The novelty and similarity of the content is higher, while the conductance and expectedness are lower. The new contract is capturing more attention, and activating different edges in the network relative to the previous segment.

    \begin{table}[bht]
    \caption{Sample Enron email stream segments}
    \begin{tabular}{ | l || c | c |}
    \hline \hline
    Parameter & Sample A & Sample B \\ \hline
     & \includegraphics[scale=0.15]{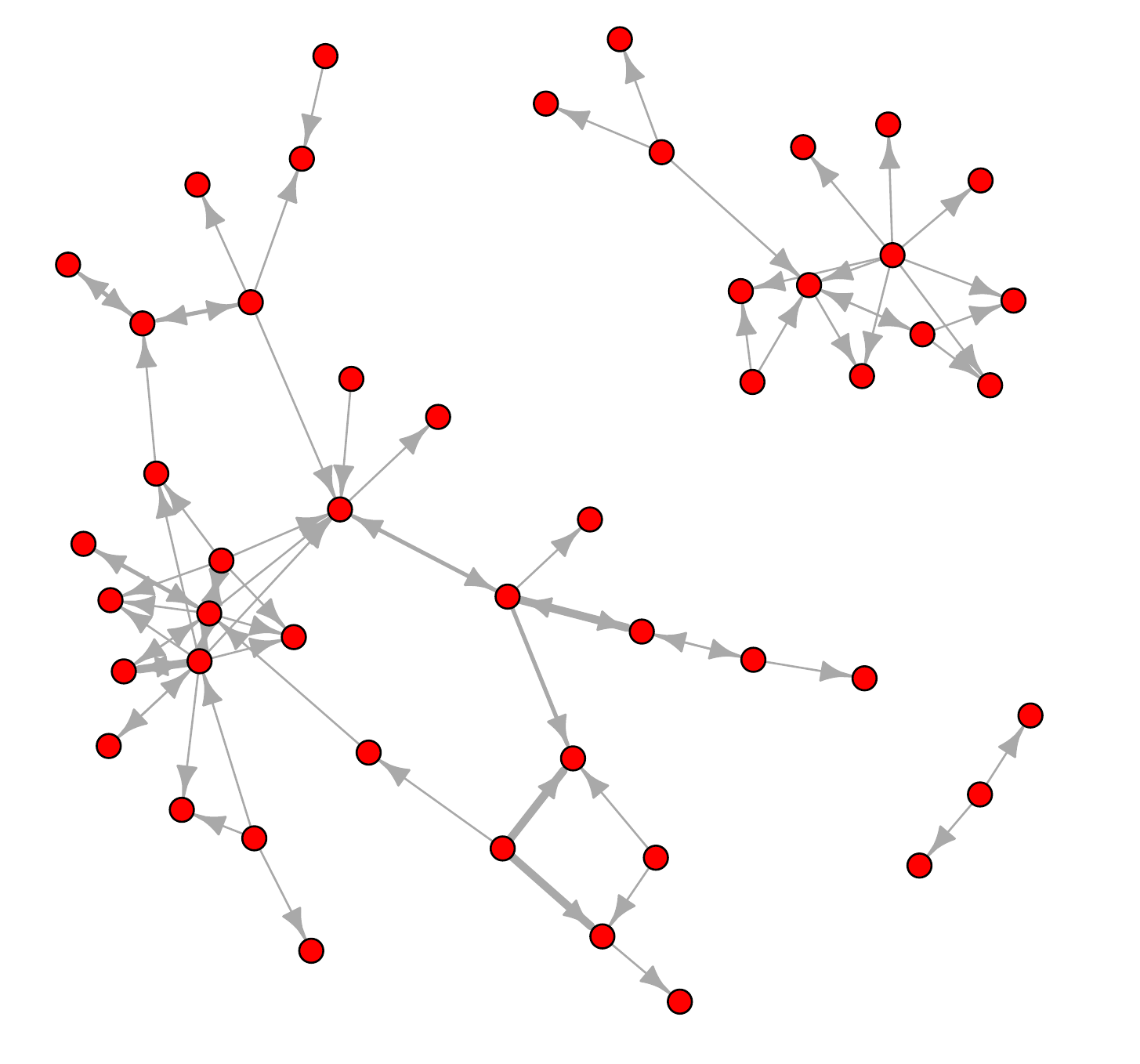} 
     &\includegraphics[scale=0.15]{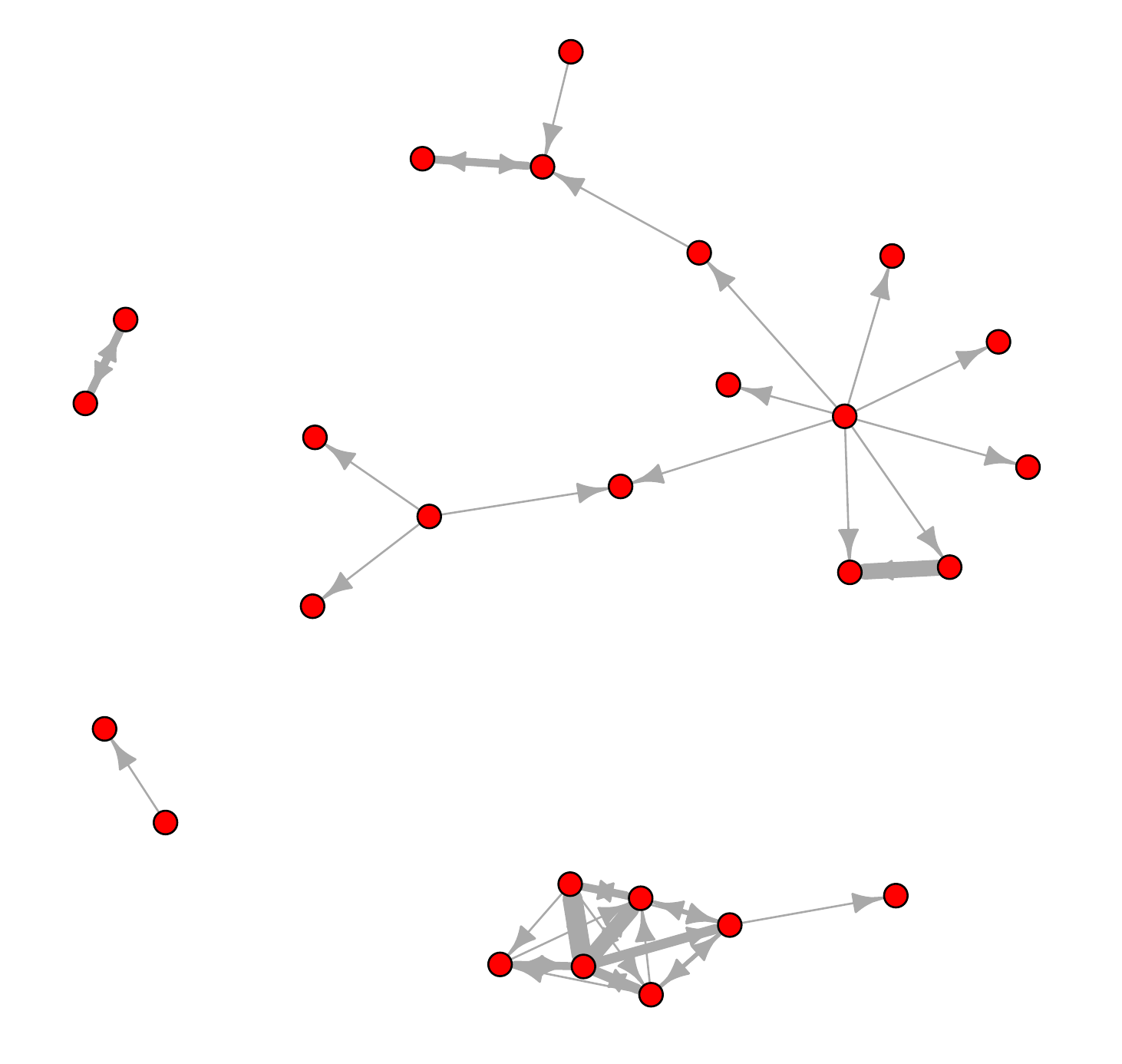} \\ \hline
   $C$ & 44.02 &  25.22 \\ \hline
   similarity &  0.049 &  0.082 \\ \hline
   $X$ & 0.410 & 0.240 \\ \hline
   LM dist. & 5.65 & 8.44 \\ \hline \hline
    \end{tabular}
    \label{table:EnronSample}
    \end{table}

\begin{figure}[bht]
\centering
\includegraphics[width=0.9\columnwidth]{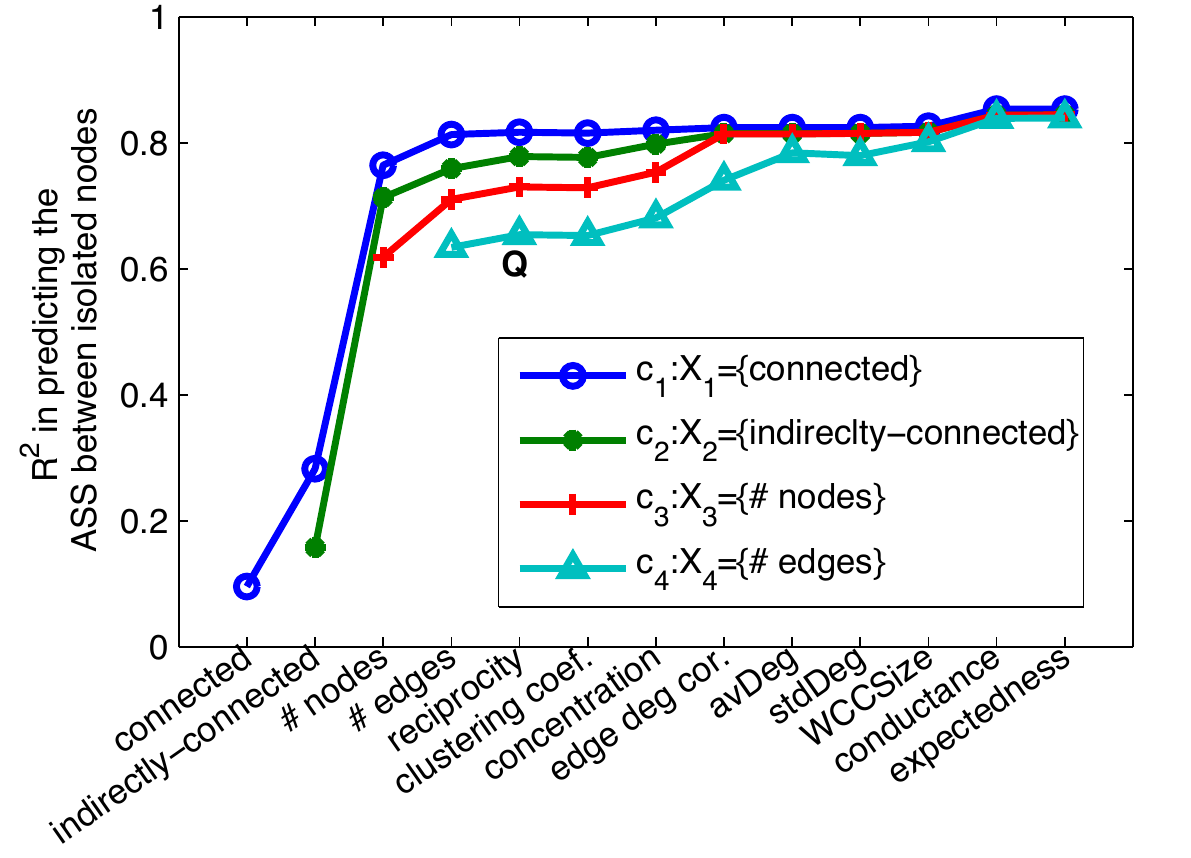}
\caption{The coefficient of determination $R^2$ of the NW kernel regression model in predicting the average semantic similarity between the tweets of non-connected nodes within a community of Twitter users. The training input set of curve $c_i$ starts from the variable $\mathcal{X}_i$ and incorporates additional variables into the model as it moves along the $x$-axis.\label{fig:tsirakis_KR}}
\end{figure}

\subsection{Network structure and content diversity} 
Next we analyze all segments together, by using a linear regression employing all network variables to predict asset entropy $H_{A}$ in Second Life. Network variables explain, across the groups, 42\% of the variance ($R^{2} = 0.42$) on average with
a standard deviation in $R^{2}$ of 0.17. The same network variables can explain 30\% of the variance in textual similarity $S(x, y)$ of Enron emails when forwarded content is excluded from the emails, and 23\% when not. 

For Twitter, we apply the Nadaraya-Watson (NW) kernel regression \cite{Nadaraya1964} to predict the content similarity which is defined as the average of cosine similarity between two nodes. For example, to infer the similarity of tweeted content between users in a Twitter community who are neither directly nor indirectly linked, one could use content variables, such as the similarity between user pairs who are linked, use just network variables, or use both.  Figure~\ref{fig:tsirakis_KR} shows the $R^2$ performance of a non-parametric regression on the series extracted from the 800-segmented tweets posted by the community shown in Figure~\ref{fig:ASS}. 

Remarkably, regressions using network variables alone perform as well as regressions incorporating content variables. 
Over all 9 Twitter communities, network variables can explain $R^{2} = 0.54 \pm 0.23$ of the variance in similarities between content broadcast by disconnected pairs, $R^{2} = 0.27 \pm 0.13$ of the variance in similarity between indirectly connected pairs, and $R^{2} = 0.18 \pm 0.15$ of similarity variance for directly connected pairs.

\begin{figure}
\centering
\includegraphics[width=0.95\columnwidth]{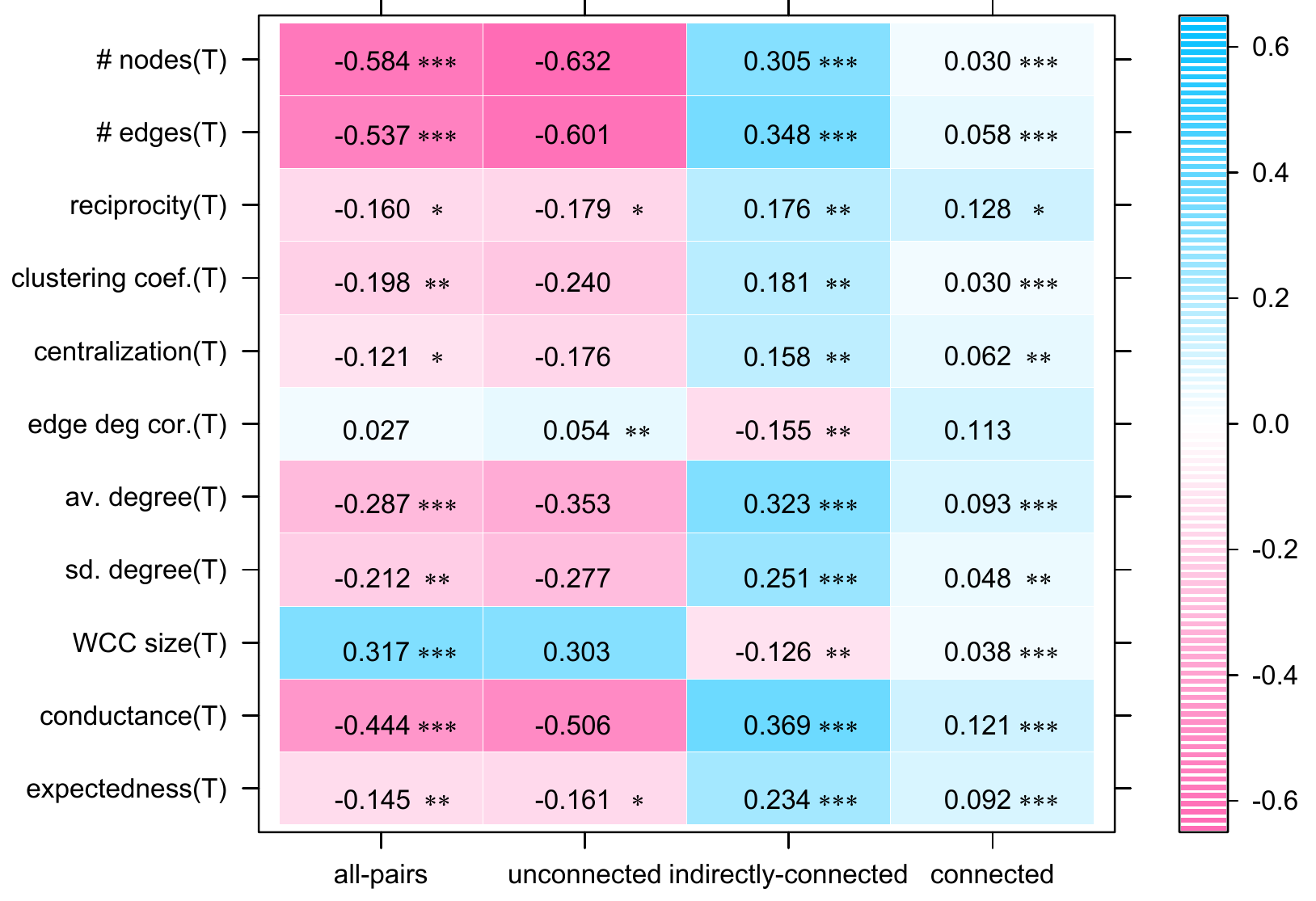}
\caption{Correlations between network topological and content variable time series averaged over 9 Twitter communities. $^{*}$ denotes $0.01 \leq p < 0.1$; $^{**}$ denotes $0.001 \leq p < 0.01$; $^{***}$ denotes $p < 0.001$. }
\label{fig:heatmap_spearman_cc}
\end{figure}

The above results indicate that network variables can be powerful indicators of how diverse the content is that is being communicated over the network. Next we elaborate on the correspondence between individual network variables and content similarity. We start with Twitter, because it shows both the strongest and most multi-faceted correlations between network and content variables, shown in Fig.~\ref{fig:heatmap_spearman_cc}. Since the number of tweets per segment is held constant, a greater number of nodes implies participation that is more distributed, while a higher number of edges corresponds to more interaction. Network segments that are larger and denser, as reflected by the number of edges, the average and standard deviation in degree, the average size of the weakly connected component, and conductance, tend to have greater semantic similarity between indirectly connected nodes, but less semantic similarity between nodes that remain in separate components. Indirectly connected nodes are also more similar when there is a greater amount of reciprocity and clustering in the network, both being indicative of social interaction. In other words, nodes that are in the same component, but not directly referring to one another, will tend to talk about shared topics if there is more participation and social interaction within the network. However, nodes that are left out during such times will tend to differ more from those they are disconnected from.

We note that network variables, with the exception of centralization and reciprocity, have little bearing on the similarity of content communicated by Twitter users who {\em directly} mention one another. That is, when nodes are directly linked, their similarity tends to be higher and relatively independent of what is occurring in the rest of the network.  Reciprocity does affect similarity for directly connected users, since two nodes reciprocally addressing one another are likely to be involved in a conversation on the same topic. Centralization does increase the likelihood that linked pairs communicate similar content.  In this case, centralization may be caused by one node driving discussion on a single topic. However, nodes that are not directly connected, meaning that they are not part of such a star formation, tend to be less similar in their discussion during segments with high centralization. 
\begin{figure}
\centering
\includegraphics[width=1.05\columnwidth]{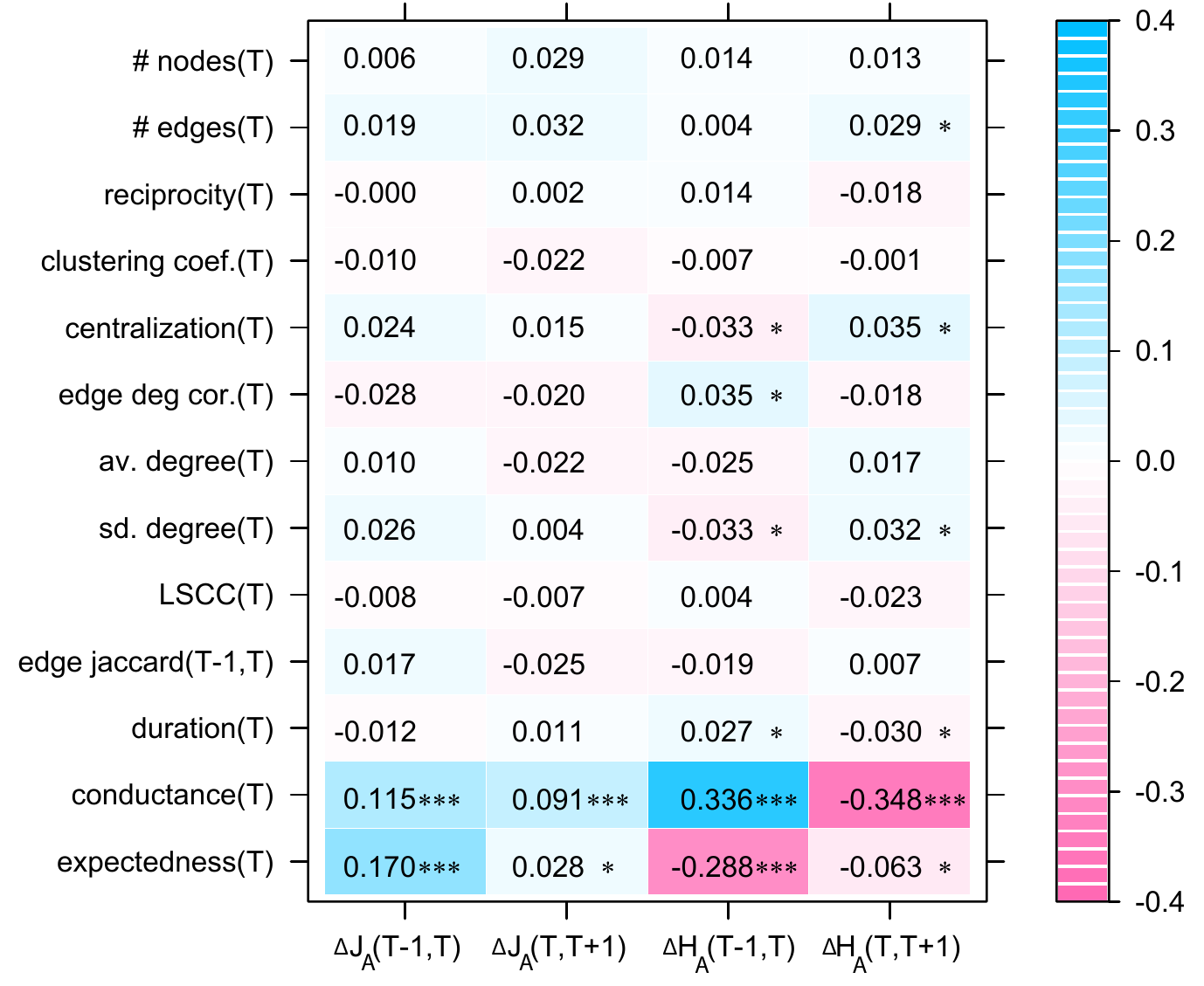}
\caption{Correlations between network topological and content variable time series averaging over 100 groups in SecondLife asset transfer graphs. }
\label{fig:heatmap_SL_tran}
\end{figure}

In contrast with Twitter communication, asset transfer networks within SecondLife groups show little to no correlation between standard network metrics and the diversity of assets being traded (see Figure~\ref{fig:heatmap_SL_tran}). Unlike in the analysis of Twitter communities, where many tweets whose content was analyzed were not associated with an edge, here all content recorded in a segment corresponds to an interaction between users. Consequently, variables, such as the number of edges, no longer capture the overall amount of interaction, but the amount of repeat interaction.  The correlation between the diversity of information and any of the standard network structural variables does not exceed 0.1. However, conductance and expectedness do show significant correlations. Conductance at time $t$ is moderately positively correlated with both the entropy of the assets being traded at $t$, and the increase in entropy from $t-1$. This means that when items are able to circulate in the network, the structure supports a higher diversity of content.

\begin{figure}[htb]
\centering
\includegraphics[width=0.97\columnwidth]{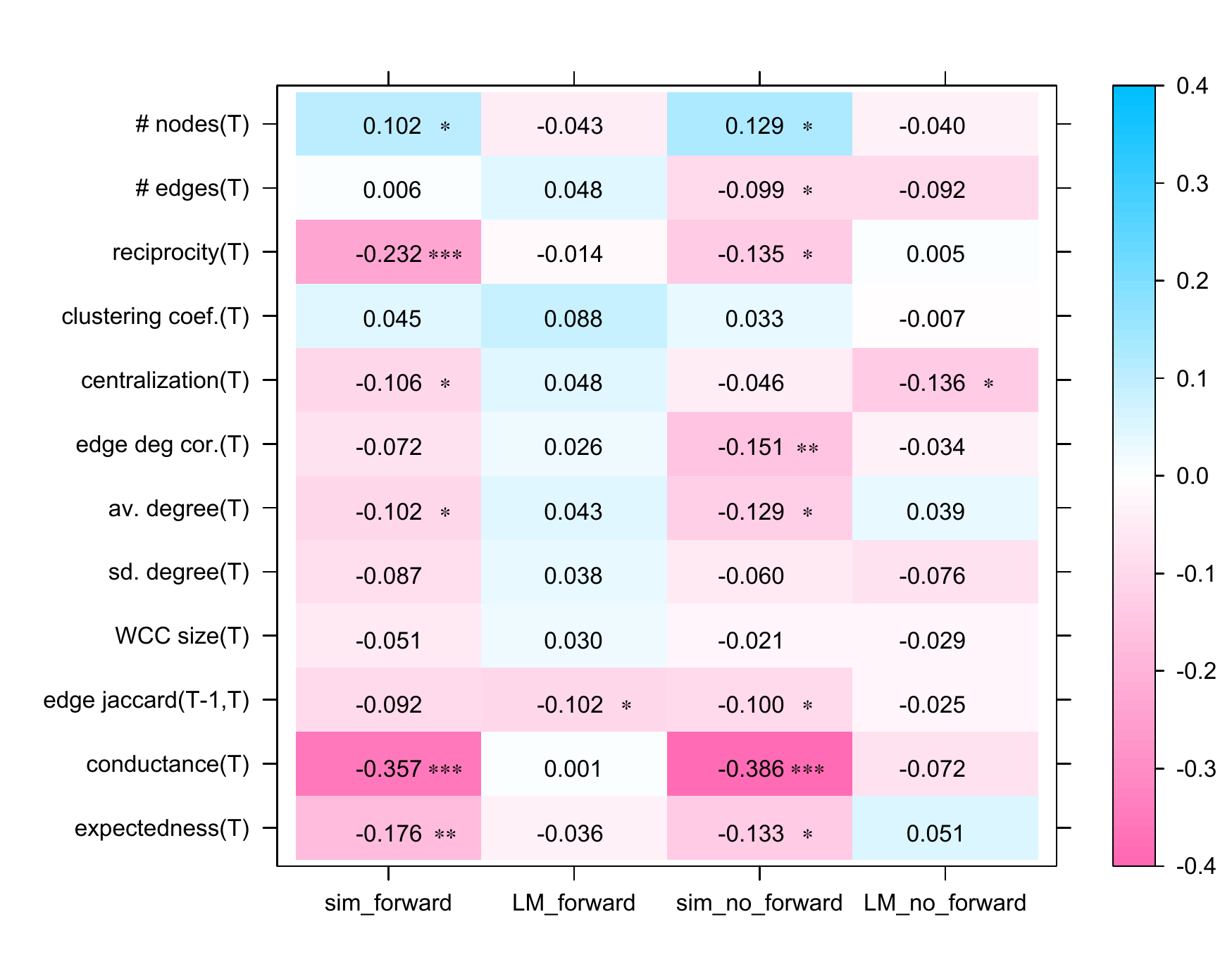}
\caption{Correlations between topological and semantic variables in the Enron email data set. {\em forward/no\_forward} denotes whether quoted and forwarded text was included in the analysis.}
\label{fig:enron}
\end{figure}

\enlargethispage{\baselineskip}
In the Enron email dataset, communication shows weak correspondence between most standard network metrics and similarity (Spearman's $\rho \leq .25$), as shown in Figure~\ref{fig:enron}. Reciprocity, clustering, and component size are slightly positively correlated with more uniform content, indicating that conversations are taking place between pairs and groups of individuals. 
Conductance is negatively correlated with content similarity ($\rho = -0.39$), consistent with both the average over all pairs in the Twitter communities, and with SecondLife groups. This result is not necessarily intuitive. High conductance implies that information can flow quickly between a large number of nodes, creating the potential for the same information to flow through a substantial portion of the network. Instead, we observe that high conductance in a network typically corresponds to a greater diversity in content.

\subsection{Network structure and content novelty} 
So far we have shown a contemporaneous correspondence between network structure and content. Next, we show that the structure can also reveal  whether the information is novel. Secondlife groups on average achieved an $R^{2}$ of 0.21 with a standard deviation of $0.15$. For Twitter, we construct two separate language models, one for all the nodes in the network and one just for non-isolated nodes, meaning that they either mentioned or were mentioned by someone (in contrast, in the SecondLife and Enron networks, we only have non-isolated nodes). For standard network metrics, which only capture the network's structure at a given timepoint, we use their 1$^\text{st}$ order derivatives to measure changes in network structure. For example, we might be interested in whether an increase in centralization, e.g. the appearance of star formations in the network, is correlated with the arrival of novel information. As with SecondLife and Enron, the novelty of content in Twitter is more difficult to predict than its similarity. Across the 9 Twitter communities, the $R^{2} = 0.19 \pm 15$ for predicting the KL divergence for non-isolated nodes, and for isolated nodes $R^{2} = 0.14 \pm 0.17$.

\begin{figure}
\centering
\includegraphics[width=0.97\columnwidth]{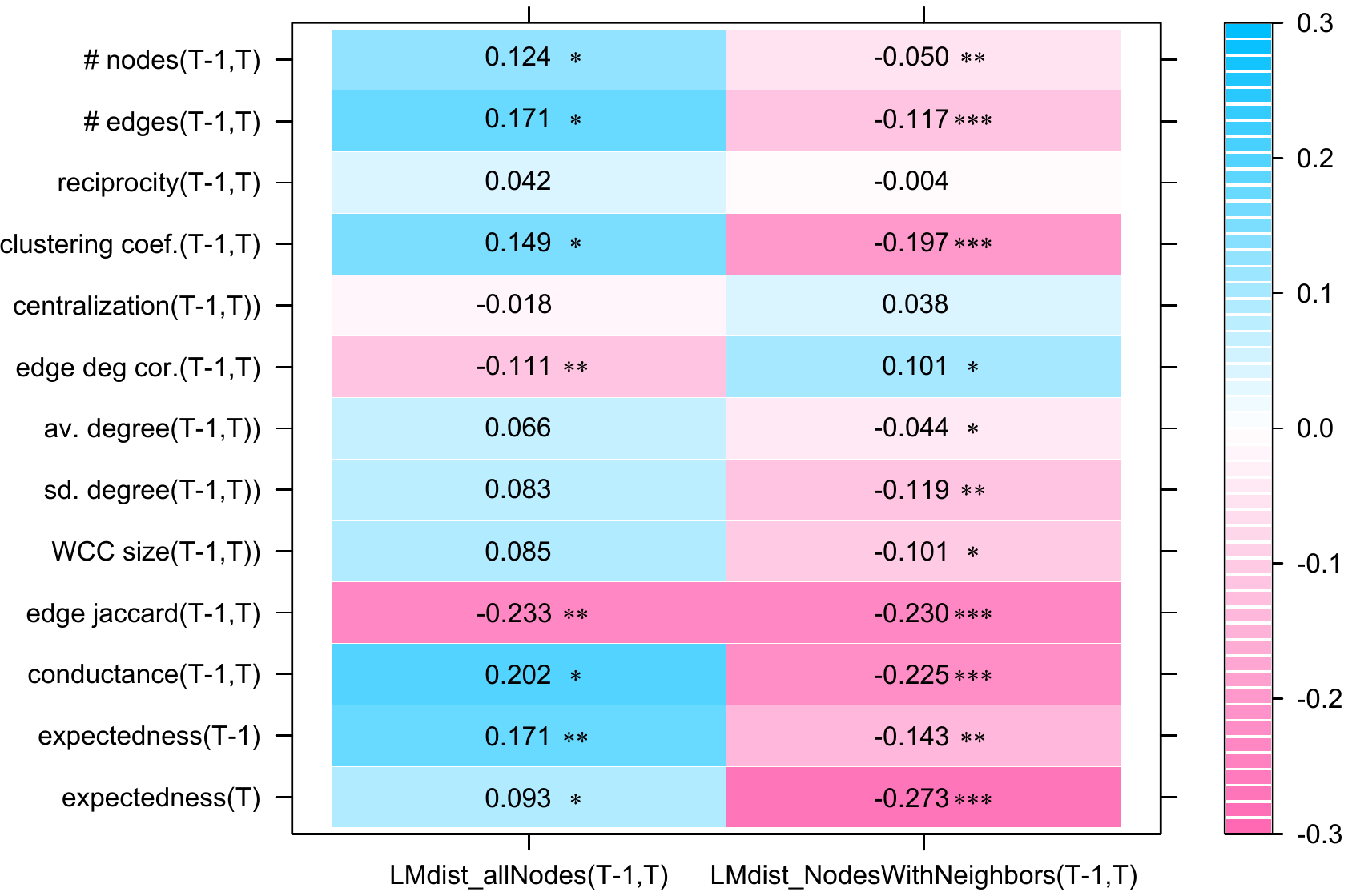}
\caption{Correlation coefficients between the LM distance metric and network structure metrics averaged over 9 Twitter communities.}
\label{fig:heatmap_LM_mean}
\end{figure}

Looking at the correlations between specific network metrics and content novelty, shown in Fig.~\ref{fig:heatmap_LM_mean}, we observe that novelty is more strongly correlated with structural network metrics of change or the potential for change, captured by the edge Jaccard $J_{E}$ and expectedness, than it is with the derivatives of standard network metrics.
Specifically, we find that if the network is relatively static, in that the overlap of edges from time period to time period is greater, then the content communicated by non-isolated nodes tends to be less novel.

SecondLife groups exhibit patterns similar to those of non-isolated nodes in Twitter, with the exception that whether or not the network is static, as captured by the edge Jaccard coefficient $J_{E}$, does not have bearing on the degree of novelty in the content. Instead, it is expectedness, which factors in the likelihood of past information flow between a newly observed edge, that is correlated with a greater sameness of content ($\rho = 0.17$). This implies that unexpected edges are more likely to occur with the arrival of new information. Additionally, when the network configuration is surprising, i.e. when expectedness is low, the information entropy tends to decrease ($\rho = -0.29$). 

Intuitively, a piece of information is stimulating new edges to form when it is novel and when it is capturing a considerable amount of attention.  That novel information within SecondLife tends to capture more attention, thus reducing diversity, is evidenced by a positive correlation ($\rho = 0.19, p < 10^{-14}$), between entropy and overlap in assets from segment to segment. 

Finally, novelty of content in Enron is not easily predicted using network metrics. As in Second Life groups, structural similarity, given by the overlap of edges $J_{E}$ between segments, is not a significant predictor of novelty. However, neither is expectedness. Nevertheless, we do find an interesting correspondence between structure and content novelty when we allowed automated log messages to remain in the Enron data set. These log mail messages comprised 18\% of all content, and were sent primarily from one email account to a list of recipients. They contained automated updates on power transmission rate schedules.  As a result of the inclusion of log messages, more uniformity of email content, boosted by the presence log messages, was less likely to be novel relative to the previous time segment ($\rho = -0.67$), in contrast with the patterns in Second Life. There was a strong negative correlation between conductance and similarity ($\rho = -0.74$) and a moderate positive correlation between expectedness and novelty ($\rho = 0.31$). 

The inclusion of computer-generated content, with its rigid form and schedule, had altered the correspondence between email network structure and content. Thus we see that the different types of correspondence between topological and semantic variables are a direct reflection of the differences in the properties of the semantic content across the different settings.

\section{Simulation model}\label{sec:simulation}
The above analysis established empirically a correspondence between the structural shape of the network and the content being transmitted, but leaves the question of how this correspondence arises and why it varies across different environments. We next   turn to a simulation to help understand directly how content influences the activation of edges and how transmission of content is influenced by the network.  

The simulation model is built on a randomly generated network of tunable size, density, reciprocity and clustering coefficient. We report results for networks of 100 nodes, 1000 edges, reciprocity=0.5, and clustering=0.3. The edges are assigned random weights to represent the closeness, i.e. the probability of passing messages between two nodes. Message content is a random Gaussian number ranging from 1 to 100. This allows us to distinguish messages, but also have a measure of similarity between them. Each node has its own message queue to keep track of the timestamp of received messages, their content, and whether it has already sent them out. 

Each node also has a policy to determine whether it will forward a message. The policy considers three message variables: recency, novelty and topicality. 
The dominant topic for the whole network is generated randomly and its lifetime is a Poisson random variable. A node calculates the novelty of a message by taking the product of the mean of the absolute difference with other messages that it has received recently. Recency decays exponentially as a function of the difference between the current timestamp and the received timestamp. Topicality  is the difference between the message and the current topic. A message is sent if the combination of these variables is above threshold. 

We collected two data sets generated from the model, one with, the other without topicality. Both data sets consist of 90,000 messages including source, destination, and content. Similar to previous sections, the simulation data is segmented by 100 messages, and novelty and diversity of content is measured per segment. The novelty of content is calculated using two variables: distance between two contiguous segments, which is defined as the sum of the absolute difference of messages between two segments, and the Jaccard coefficient, which measures the proportion of the overlap in messages between two contiguous segments. The diversity of content is measured by the entropy of the message distribution and the standard deviation of message values per segment. 

\begin{figure}[h]
\centering
\includegraphics[width=0.97\columnwidth]{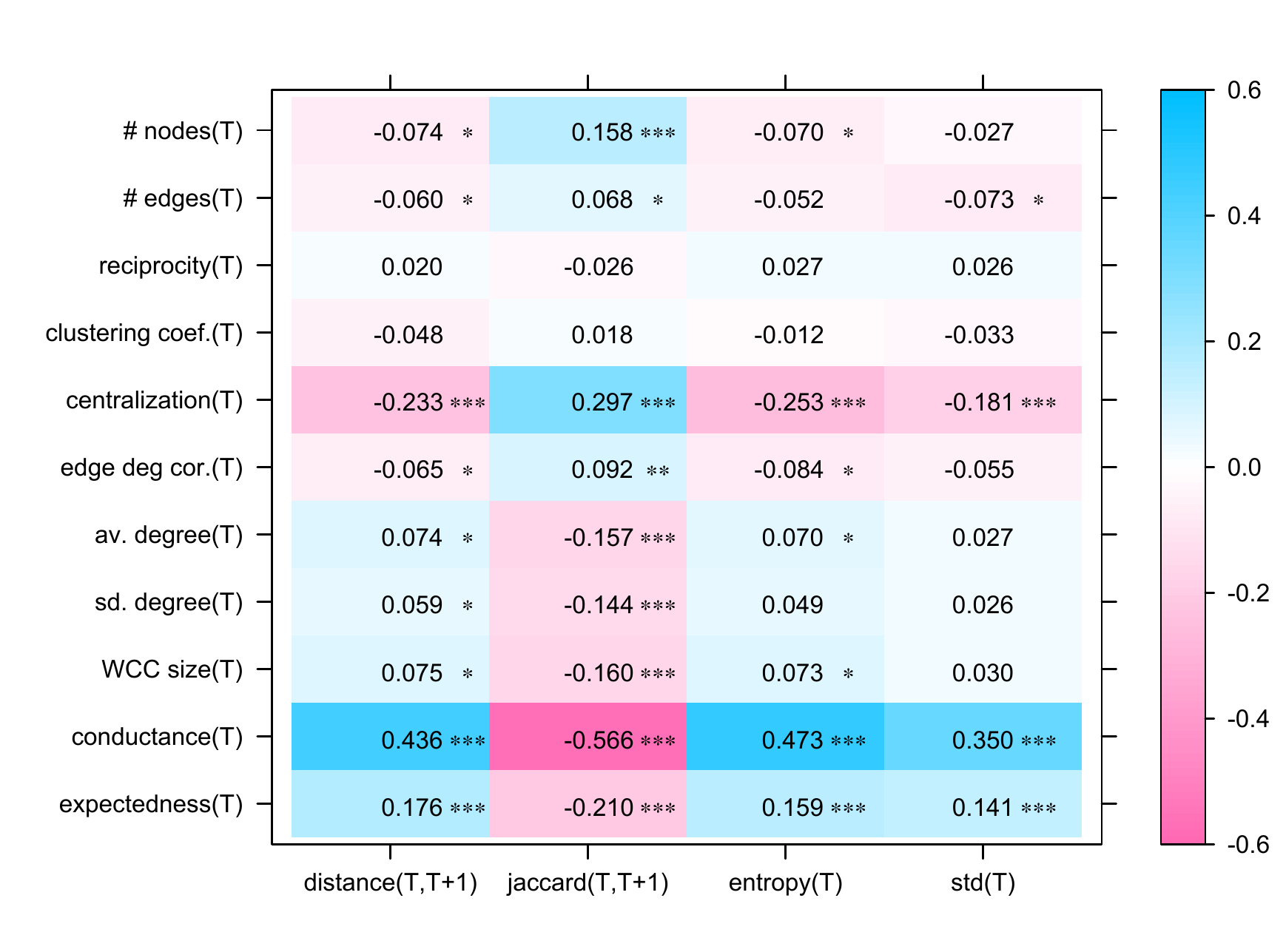}
\caption{Correlations between topological and semantic variables in simulated data without topicality.}
\label{fig:heatmap_notopic}
\end{figure}

The model without topicality (see Figure ~\ref{fig:heatmap_notopic}) shows patterns similar to the Twitter dataset including isolated nodes and edges: high conductance corresponds to greater diversity in content, while higher expectedness in network structure corresponds to greater novelty. This result implies that Twitter users are not generally driven by a general topic but by diverse topics, and this finding is consistent with Ryan's Twitter study \cite{kelly2009twitter} which shows that most of tweets are uninformative, and only $8\%$ of tweets have pass-along value. 

\begin{figure}[h]
\centering
\includegraphics[width=0.97\columnwidth]{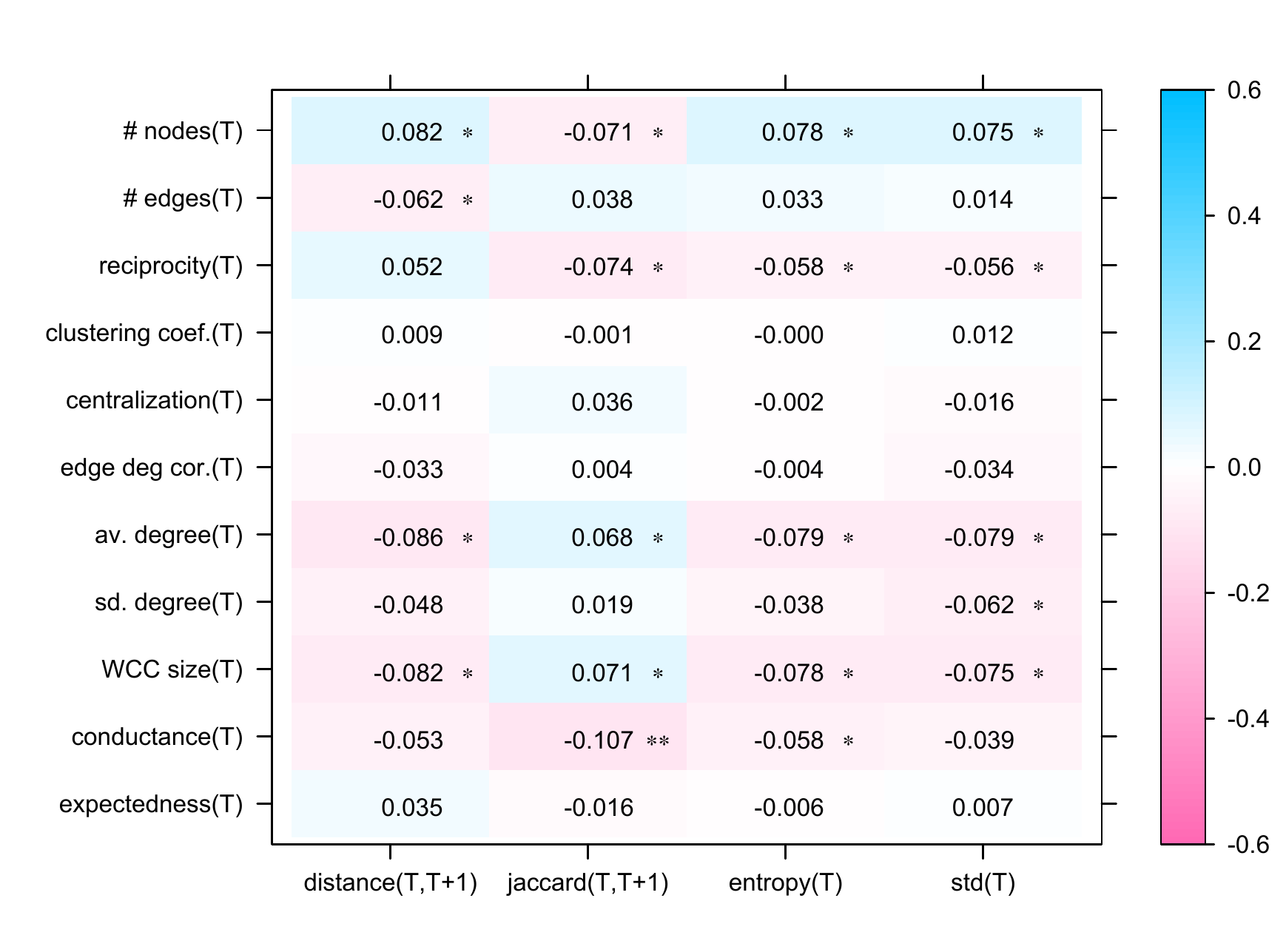}
\caption{Correlations between topological and semantic variables in simulation data with topicality. }
\label{fig:heatmap_topic}
\end{figure}

Once topicality is included (Figure ~\ref{fig:heatmap_topic}), the correspondence between network and content variables is altered. We find that the inclusion of topicality reduces the diversity of content, measured as the average message entropy (2.47 vs. 2.66) and standard deviation (13.8 vs. 19.3). Novelty is also depressed with the inclusion of topicality. The message Jaccard coefficient rises from $\mu = 0.12$ to $\mu = 0.71$, and the message distance drops from$\mu = 24.18$ to $\mu = 23.35$.  The inclusion of topics also reduces the correspondence between novelty and diversity of content. For instance, the correlations between the Jaccard coefficient and the entropy weaken  (from $-0.64$ to $-0.3$, $p < 0.001$) 
with topics. 
As expected, topicality keeps the nodes talking about the same thing for a longer period.

The patterns exhibited by topic-driven simulations are similar to what we observe for Enron, SecondLife, and Twitter data without isolated nodes. This finding suggests that the communication within these networks is topic-driven. One may speculate that the general topic of Enron is driven by the operation of Enron company, and SecondLife information exchange is driven by events and interests.  Twitter users connected by tweets often share interests in a common topic. 

\section{Discussion}\label{sec:conclusion}
The above findings indicate that much information can be gleaned by considering the evolution as opposed to the static configuration of a network.  In three diverse online settings, we tied the evolution of network structure to coherence and novelty of content, being able to explain a substantial portion of the variance in content variables using network variables alone. 

Beyond establishing an expected correspondence between structure and content, the analysis yielded two interesting insights. The first is that the most obvious network metrics, such as the number of nodes and edges, reflecting overall activity, are not always the most informative. In two of the settings, Second Life and Enron, these metrics did not net significant information about the content. However, two measures did provide insight. Conductance, which captures the potential for information exchange, was correlated with high and increased diversity of information content. A similarly derived measure of network expectedness was correlated with content novelty. 

The second insight is that the relationship between network and content variables is sensitive to the way the content itself evolves and is exchanged within the system. For example, in SecondLife trades, the appearance of novel assets tends to correspond to reduced diversity. Novel information captures more attention. In contrast, diversity and novelty were uncorrelated in content emailed in a corporate setting. This was clarified in the simulation model, which showed that the tension between novelty and topicality can produce patterns of communication behavior matching the different data sets. 

However the exact relationship between network structure and communicated content may vary from system to system, the correspondence is high enough for network variables to be able to explain a significant portion of the variance in content novelty and diversity. Thus the results point to the feasibility of inferring the properties of information flowing over the network from the structure of the network alone. 

Another factor is the manner in which the network is constructed.  For instance, we aggregated all transmitted content at the user level in Twitter due to the sparsity and brevity of directed communication. In this case, most network variables could be used in predicting content properties, since the network captured how much of the communicated content was in connection with other nodes within the community. This points to the need to calibrate the correspondence between network and content variables to a specific setting before one can start predicting content properties from network structure.

In the context of the web, the technique could be used to help allocate attention and resources to a particular portion of it. Is there signal in the network structure revealing that there is content being transmitted that is worth paying attention to? Another application would be to track and facilitate response to small and large-scale disasters as it is reflected in social media or the communication patterns of emergency response personnel. Outside the context of the Web, the technique could be useful in scenarios where content cannot be easily and continuously accessed, but the network structure of communication is easily observable. Such situations may arise in monitoring adversarial networks, where intercepting or deciphering information is costly, but the communication frequency, as well as sources and targets, may be observable. 

In future work we intend to apply these methods to understanding the changes in the co-evolution of structure and content, by differentiating periods of bursty activity~\cite{kleinberg2003bursty} from normal rates of change. We will examine whether network and content change tends to be instigated by a few nodes consistently, or whether the changes observed are an emergent property of the system. We expect that this kind of analysis will not only be useful in inferring information content properties, but also in characterizing the fitness of the network itself to transmit and adapt to information in a dynamic way.

\section{Acknowledgements}
We would like to thank Eytan Adar and Mark E. J. Newman for helpful discussions. The work was supported in part by NSF IIS-0746646, U.S. Air Force Office of Scientific Research MURI award FA9550-08-1-0265, and by the U.S. Army Research Laboratory under Cooperative Agreement No. W911NF-09-2-0053 (NS-CTA). The views and conclusions contained in this document are those of the authors and should not be interpreted as representing the official policies, either ex- pressed or implied, of the U.S. Government. The U.S. Government is authorized to reproduce and distribute reprints for Government purposes notwithstanding any copyright notation here on.    

\small
\vspace{4pt}
\bibliographystyle{acm-sigchi}
\bibliography{NetCoevolutionGong}

\end{document}